\documentclass[preprint,authoryear]{elsarticle}

\makeatletter
\def\ps@pprintTitle{
    \let\@oddhead\@empty
    \let\@evenhead\@empty
    \def\@oddfoot{\footnotesize\itshape
    {Accepted Manuscript in Icarus, 2023. \url{https://doi.org/10.1016/j.icarus.2023.115843}. Shared here under CC-BY-NC-ND 4.0.}\hfill}
    \let\@evenfoot\@oddfoot
    }
\makeatother

\journal{Icarus}
\usepackage{geometry}
\usepackage{amsmath}
\usepackage{amssymb}
\usepackage{amsfonts}

\usepackage{siunitx}

\usepackage{enumitem}
\setlist[itemize]{noitemsep, topsep=0pt}
\setlist[enumerate]{noitemsep, topsep=0pt}

\usepackage[hidelinks]{hyperref}

\usepackage{float}
\usepackage[font=small,skip=0pt]{caption}

\usepackage{dsfont}
\usepackage{graphicx}

\usepackage{algorithm}
\usepackage{algorithmic}

\usepackage{microtype}

\newcommand{\pois}{\text{Poisson}}

\newcommand{\B}[1]{\mathbf{#1}}

\date{\vspace{-3em}}

\begin{document}

\begin{frontmatter}
\title{GOES GLM, Biased Bolides, and Debiased Distributions}

\author[1]{Anthony Ozerov\corref{cor1}}\ead{ozerov@berkeley.edu}
\author[2,3]{Jeffrey C. Smith}
\author[3]{Jessie L. Dotson}
\author[4]{Randolph S. Longenbaugh}
\author[2,3]{Robert L. Morris}
\cortext[cor1]{Corresponding author}

\affiliation[1]{organization={Department of Statistics, University of California, Berkeley}, city={Berkeley}, postcode={94720}, state={California}, country={United States}}
\affiliation[2]{organization={SETI Institute}, city={Mountain View}, postcode={94043}, state={California}, country={United States}}
\affiliation[3]{organization={NASA Ames Research Center}, city={Moffett Field}, postcode={94043}, state={California}, country={United States}}
\affiliation[4]{organization={Sandia National Laboratories}, city={Albuquerque}, postcode={87123}, state={New Mexico}, country={United States}}


\begin{abstract}
\noindent The large combined field of view of the Geostationary Lightning Mapper (GLM) instruments onboard the GOES weather satellites makes them useful for studying the population of other atmospheric phenomena, such as bolides. Being a lightning mapper, GLM has many detection biases when applied to non-lightning and these systematics must be studied and properly accounted for before precise measurements of bolide flux can be ascertained. We developed a Bayesian Poisson regression model which simultaneously estimates instrumental biases and our statistic of principal interest: the latitudinal variation of bolide flux. We find that the estimated bias due to the angle of incident light upon the instrument corresponds roughly with the known sensitivity of the GLM instruments. We compare our latitudinal flux variation estimates to existing theoretical models and find our estimates consistent with GLM being strongly biased towards high-velocity bolides.
\end{abstract}

\end{frontmatter}

\section{Introduction}

\noindent Asteroids and meteoroids do not necessarily impact the Earth uniformly \citep{darrel}. Based on the modeled distribution of orbits of Near-Earth Objects (NEOs) and the corresponding long-term possible impact configurations and probabilities, the distribution of impacts across latitudes can be estimated. This has been done by \cite{lefeuvre}, with \cite{darrel} taking the methods further by taking into account a broader NEO population and orbital precession effects. In terms of empirically verifying the estimates, the two main methods used in the literature are cratering records and impact detections.
Cratering records are applicable to bodies like the Moon \citep{darrel, lefeuvre}, but are difficult to obtain on the Earth due to weathering. For impact detections, there are several possible sources of data.

Previous work has used impact detections from \textbf{United States Government} (USG) sensors \citep{darrel,evatt}. This set of data is appealing as it has global coverage. However, due to the classified nature of the sensors, the detection biases of this data set are unknown to the scientific community. Because of the defense-related priorities of these sensors and unpublished release schedule for bolides, the published dataset may not necessarily represent a uniform detection efficiency across different latitudes. For instance, the sensors might generally have a lower detection efficiency for objects hitting the poles (which would explain all of the discrepancy described in \cite{darrel}). This makes it impossible to use this data to probe the distribution of bolide impacts across latitudes.

\textbf{Ground-based systems} also detect bolides. Automated ground-based photo, video, and radar meteor observation stations detect thousands of meteor impacts daily. These include the Global Meteor Network \citep{gmn}, Cameras for Allsky Meteor Surveillance \citep{cams}, the Canadian Meteor Orbit Radar \citep{cmor}, and many others. Yet they mostly detect small objects from known meteor showers. Their limited coverage in area and consequent small area-time product makes it more difficult to study the distribution of rarer, larger objects of greater concern for planetary defense, simply due to insufficient data. This limited coverage is also still a problem for large dedicated ground-based systems for detecting bolides (as opposed to relatively dim meteors), like the Desert Fireball Network \citep{hadrien} and the Fireball Recovery and InterPlanetary Observation Network \citep{fripon}. Moreover, some systems are unable to detect objects during the day, and there are other factors like weather and the specifics of a given system that may nonuniformly affect detection efficiency across different latitudes.

Finally, the \textbf{Geostationary Lightning Mapper} (GLM) instruments \citep{goodmanglm} onboard the Geostationary Operational Environmental Satellites (GOES) are known to detect bolides \citep{ogglm,rumpfglm}. This data source is appealing as it has near-hemispheric coverage and much more is publicly known about GLM than USG sensors. In previous studies, like \cite{darrel}, only human-vetted data, which has a bias towards the stereo region and highly varied detection methodology over time, has been used.
Other biases, like the angle of incident light upon the sensor, have not previously been accounted for. This work seeks to account for the biases in GLM data and produce a more accurate estimate of the variation in bolide impacts across latitudes.

\begin{figure*}
    \begin{center}
    \includegraphics[width=0.9\textwidth]{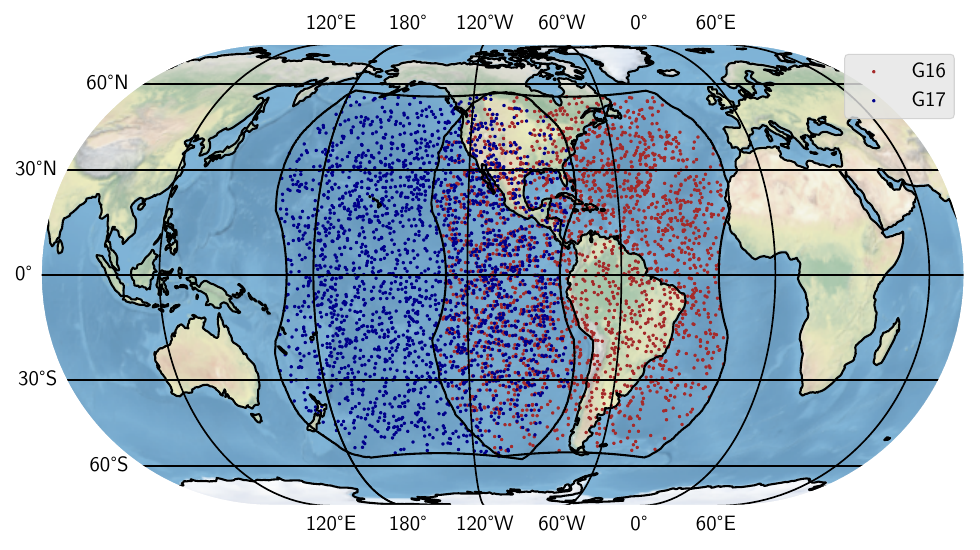}\end{center}
    \caption{Detections with confidence $>0.7$ in the \texttt{machine} database used, colored by which satellite detected them. There are double-detections in the stereo region, where the same object was detected by both satellites. The offset between these double-detections is due to parallax. As the detection location is determined according to the fixed GLM lightning ellipsoid while bolides typically occur higher in the atmosphere, due to parallax GOES-17 detections in the stereo region are consistently East of their GOES-16 counterparts. The boundaries shown are the two platforms' FOV boundaries, which closely match the limits of the bolide detections. Eckert IV projection.}
    \label{fig:detections}
\end{figure*}
\section{Using GLM for bolide population statistics}
\noindent As GLM was not designed to detect bolides, there are several considerations when attempting to use GLM data to learn about the population of bolides:
\begin{itemize}
    \item Which set of GLM bolide detections should we use?
    \item Are there any detection biases which affect our results? How do we adjust for them?
    \item What population of objects is GLM mostly seeing? Does it coincide with the population we are attempting to make claims about?
    \item What do theoretical models say the population of GLM bolide detections should look like?
\end{itemize}
We address these points below.

\subsection{The data}

\noindent We use two different bolide databases in this study: a \texttt{machine} database and a \texttt{human} database. The \texttt{machine} database represents detections before being vetted by a human, and is shown in Figure \ref{fig:detections}, while the \texttt{human} database represents human-vetted detections.

Both databases originate from the GLM bolide detection pipeline described in \cite{jeff}. Broadly, the detection pipeline reads all Level 2 GLM group data, clusters the groups in space and time, and feeds these clusters through a trained classification algorithm. The classifier was trained on a smaller set of human-labeled data, and it outputs a confidence score for any new clusters. This confidence score is \textit{not} a probability of being a bolide. It is a number between zero and one which the classifier outputs using a function it has learned from the training data, and ideally the number is close to one for any true bolides and close to zero for all other clusters. Any clusters with a confidence score above a low cutoff value are sent to a human for vetting. Each cluster comes from only one GOES satellite but, when vetting, if there are clusters detected by the other satellite which could match it, they are displayed too along with an altitude estimate supposing that the detections in both satellites were generated by the same source. The human also sees several metrics for the cluster and images of the surrounding area taken by the Advanced Baseline Imager. More information on what the human vetter sees is available in \cite{jeff}. Those clusters which pass human vetting are published at \url{https://neo-bolide.ndc.nasa.gov}. The \texttt{human} database is all published bolide detections between 2019-07-01 and 2022-06-03.

\begin{figure}
    \begin{center}
    \includegraphics[width=0.9\textwidth]{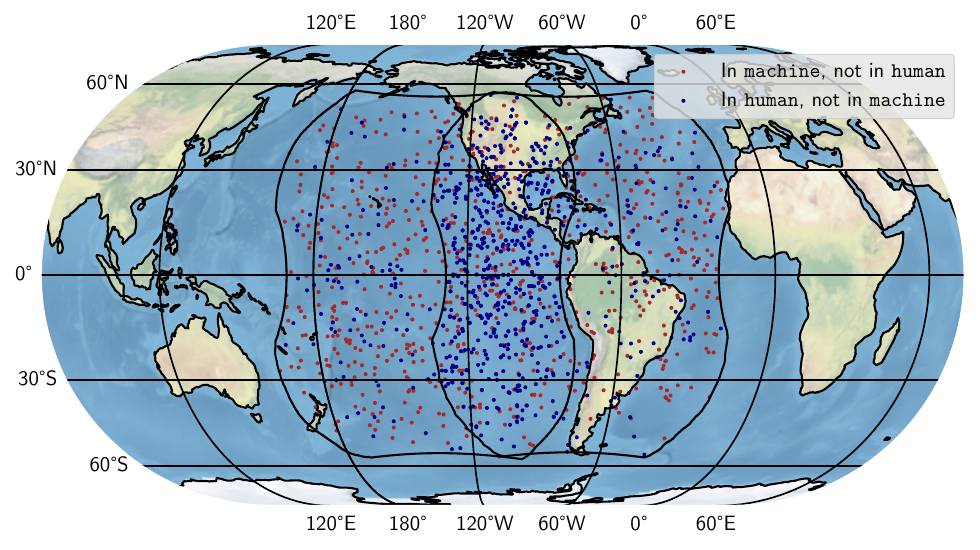}\end{center}
    \caption{Detections which appear in one database and not the other. Neither can truly be called ``false positives'' or ``false negatives,'' as some detections in $\texttt{machine}$ may be true bolides accepted by the classifier with confidence $\geq 0.7$ but rejected or never seen by a human vetter, and vice-versa. The detections that appear in $\texttt{machine}$ and not in $\texttt{human}$ are fairly evenly distributed across the field of view, and there aren't any alarming large-scale trends. The detections that appear in $\texttt{human}$ and not in $\texttt{machine}$ are concentrated in the stereo region, reflecting a human selection bias towards detections in the stereo region. Eckert IV projection.}
    \label{fig:discrepancies}
\end{figure}

Importantly, the automated part of the detection pipeline has been periodically updated. Some of these changes are general algorithm improvements, and some are simply classifier retrainings---the increasing amount of human-labeled data means that the classifier can be retrained for better performance. Updated versions of the automated portion of the pipeline can be re-applied to all historical GLM data, and this is how we obtain the \texttt{machine} database. \texttt{machine} is the set of bolide detections between 2019-07-01 and 2022-06-03 which, where not specified otherwise, contains only detections to which the classifier assigned a confidence of $\geq 0.7$. At that threshold the bolides which appear in \texttt{machine} and not in \texttt{human} are evenly distributed over the field of view, as shown in Figure \ref{fig:discrepancies}. These detections may be false positives, but they could also be legitimate bolides which were never seen by a human vetter. Reprocessings of historical GLM data have never been vetted by humans due to labor constraints. Any legitimate bolides in \texttt{machine} which were not picked up by early versions of the automated detection pipeline were never seen by a human.

We also use versions of \texttt{machine} with other thresholds, and find that glint and lightning activity introduce a bias when far lower thresholds are chosen. The goal of applying our methods to both human-vetted and unvetted data at multiple thresholds is to confirm that our results are stable against the selection effects of the confidence threshold we choose and any selection biases of a human vetter.

The detection pipeline uses Level 2 GLM data instead of Level 0 data. While the processing done by NOAA is designed for lightning and can corrupt bolide light curves generated from Level 2 data \citep{rob}, we do not expect this processing to have a spatially heterogeneous selection effect for bolides.

In the remaining analyses in this section we deal only with the \texttt{machine} database. In our primary analysis in Section \ref{sec:methods}, we work with each database separately under the assumption that each only contains true positives. The detection pipeline also runs separately for each satellite, i.e.~each satellite makes its detections independently. Because of this, in our analysis, double-detections in the stereo region are counted twice---once for each satellite. This is important when treating each satellite as an independent detector.

\begin{table}
    \centering
    \caption{Correspondences of bolide detections between \texttt{machine} and \texttt{human}. As there is no dataset of every single true bolide within the GLM FOV, it is not possible to say exactly how many bolides appear in neither \texttt{machine} nor \texttt{human}.}
    \begin{tabular}{c|cc}
                          & In \texttt{machine} & Not in \texttt{machine}\\\hline
    In \texttt{human}     & 3909 & 926\\
    Not in \texttt{human} & 772  & ?
\end{tabular}
\end{table}

\subsection{Human selection effect}\label{sec:human}

\noindent The goal of using the unvetted pipeline data in the \texttt{machine} database is to avoid any human selection biases. In particular, a human vetter may over-emphasize the stereo region because a human becomes much more confident that an object is a bolide when it is detected in both satellites and the projected measured altitude is high above cloud-level. Conversely, if a detection is only made by one satellite when the other ought to have detected it, a human might become less confident. Additionally, if a bolide is detected in both satellites, but only one obtained a good, bolide-like light curve, both satellites will count as having detected a real bolide by virtue of the good light curve. As the combination of all of these effects, an increase in detection efficiency in the stereo region is qualitatively apparent in Figure \ref{fig:discrepancies}. In Section \ref{sec:methods}, we demonstrate how to take this effect into account when working with human-vetted data. Quantitative estimates of the effect are presented in Figure \ref{fig:human-lat} in Section \ref{sec:results}.

These human selection effects are, of course, sensible for making a database of objects very likely to be bolides (e.g.~\url{https://neo-bolide.ndc.nasa.gov}). But they also affect the geographic distribution of detections. Due to the shape of the stereo region---which, as seen in Figures \ref{fig:detections} and \ref{fig:discrepancies}, varies in its size relative to the total GLM field of view (FOV) across latitudes---previous attempts to use GLM data to study the latitudinal distribution of impacts likely exhibit artifacts of this human bias. We do note that even though the \texttt{machine} data is unvetted, the underlying classifier is still trained on human-vetted data and may exhibit some bias as a result of this.

\subsection{How to empirically obtain the latitudinal variation}\label{sec:how}

\begin{figure*}
    \centering
    \includegraphics[width=0.49\textwidth]{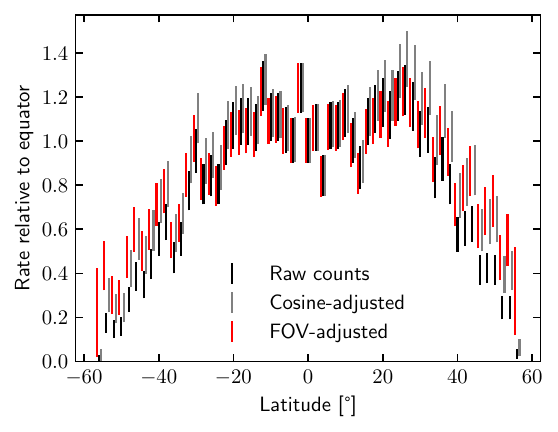}
    \includegraphics[width=0.48\textwidth]{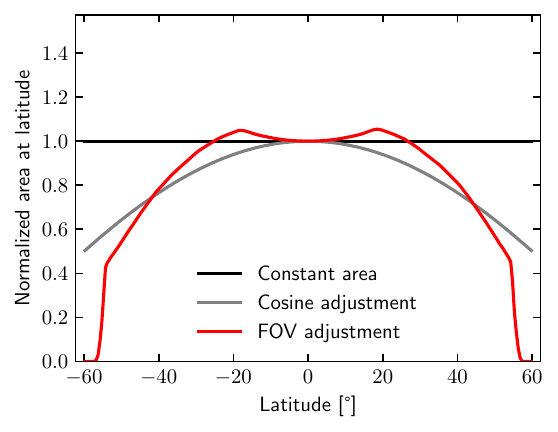}
    \caption{Left: GLM bolide detection counts split into $2^\circ$ latitude bands and normalized so that the count at the equator is 1. The errors are $1\sigma$ confidence intervals on the rates of the Poisson distributions generating the counts (these are not exactly $\sqrt{N}$ Poisson errors). Three different measurements are shown: the raw counts in each bin; the counts adjusted by the cosine of the bin's latitude; the counts adjusted by the sum of the GOES-16 and GOES-17 field of view area (\unit{\km^2}) within that bin. If bolides impact uniformly across latitudes and GLM has no detection biases, we would expect the FOV-adjusted rates to be uniform at 1. Right: The normalized size of the GLM FOV at different latitudes, compared to a uniform area and a cosine adjustment. Counts can be divided by these adjustments to correct the counts to a flux estimate at different latitudes, as is done in the left figure. Moving away from the equator, the area observed by GLM increases until about $20^\circ$, then decreases rapidly. Even adjusting for this rapid decrease in observing area does not lift the rate at the extreme edges of the left figure to parity with the rate at the equator.}
    \label{fig:lat-hist}
\end{figure*}

\begin{figure*}
    \centering
    \includegraphics[width=0.35\textwidth,page=5]{paper-diagrams}\hfill
    \includegraphics[width=0.5\textwidth]{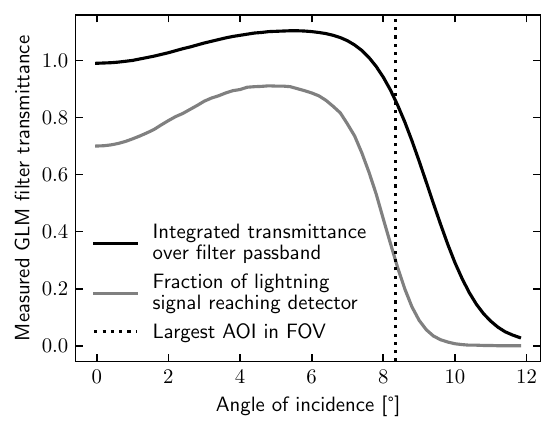}
    \caption{Left: Diagram of GOES GLM displaying the angle which we refer to as the ``angle of incidence.'' For clarity, the image used excludes the baffle which extends from the top of the instrument. Image adapted from NOAA with photo credit to Lockheed Martin (\cite{glmpic}). Right: Integrated GLM transmittance and the fraction of lightning signal reaching the detector as a function of the angle of incidence. Both curves of sensor data were measured from the GLM filter system \citep{glmfilter}. Both curves peak at around $5^\circ$ and drop off considerably at higher angles of incidence.}\label{fig:aoi}
\end{figure*}

\noindent One simple way to analyze the latitudinal variation of bolide detections in GLM data is to make a histogram of detections by latitude. This analysis is done in \cite{darrel} and is repeated here in Figure \ref{fig:lat-hist}. Looking at the raw counts, there is an observed increase in detections at about $20^\circ$ from the equator, followed by a steep drop-off. Figure \ref{fig:lat-hist} also demonstrates how important it is to correctly adjust for the area of the field of view at different latitudes. But even correctly performing this adjustment only reduces, and does not eliminate, the drop-off away from the equator. Neither does the field of view adjustment eliminate the double-peaked shape of the histogram. This analysis raises some questions:

\begin{itemize}
    \item Given the theoretical results of recent works like \cite{darrel}, these trends are unexpected, as we would expect bolide rates to increase as one moves away from the equator. Are the features of the histogram in Figure \ref{fig:lat-hist} the result of instrumental biases, or are they real?
    \item When making such a histogram, the error bars are quite large, because the bins are treated as independent. Can we relax this assumption and get tighter estimates?
\end{itemize}

\noindent In terms of instrumental biases, an important one is the angle of incidence at which light enters the aperture, which is illustrated in Figure \ref{fig:aoi}. This has been discussed before as a factor which reduces the amount of light reaching the CCD (the ``transmittance'') from a bolide as the angle of incidence increases \citep{ogglm,rob}. Both \cite{ogglm} and \cite{rob} present schematic illustrations of how a changing angle of incidence affects the effective filter passband and hence changes the amount of bolide light that reaches the CCD. It is natural, therefore, to think that the detection efficiency itself will be different at different angles of incidence, though this has not yet been studied. The light from bolides at latitudes as extreme as $\pm 50^\circ$ will, on average, have a higher angle of incidence into the aperture than that from bolides near the equator. Hence bolides at extreme latitudes can be expected to have a lower probability of detection. This leads to the possibility that perhaps the drop-off observed in Figure \ref{fig:lat-hist} is explained completely by such an effect, and thus it is a factor we account for in Section \ref{sec:methods}.

Other than the human selection effect in the stereo region discussed in Section \ref{sec:human}, we consider three additional possible detection biases that are based on properties of the bolide impact location, and not properties of the bolide itself:
\begin{itemize}
    \item Lightning flash density. In areas with high lightning flash density (expressed as lightning flashes per unit area per unit time), we may expect bolide detection efficiency to be lower. This is because the classification algorithm used contains variables correlated with locations on the Earth as predictors \citep{jeff}. We can thus suppose that it will penalize detections in areas with high lightning flash density, as otherwise there would be many false positives.
    \item Cloud cover. As clouds show up brightly in GLM, we can guess that areas which are often cloudy might have fewer bolide detections as they will, on average, have to surpass a higher noise threshold \citep{edgington2019}.
    \item Land. Land provides a different background for bolide detections than water does. It tends to be brighter, and could be more or less variable. It could also contain more or fewer false positives due to solar glint. Together, effects like these may make it easier or more difficult for GLM to detect bolides over land than over water.
\end{itemize}
Evidently, we do not know how these three factors affect GLM's ability to detect bolides. But they can all be included in our models, and the effects can be estimated. This is done in Sections \ref{sec:methods} and \ref{sec:results}, respectively, with estimates of these three possible detection biases presented in Figure \ref{fig:biases}.

\begin{figure*}
        \includegraphics[width=\textwidth]{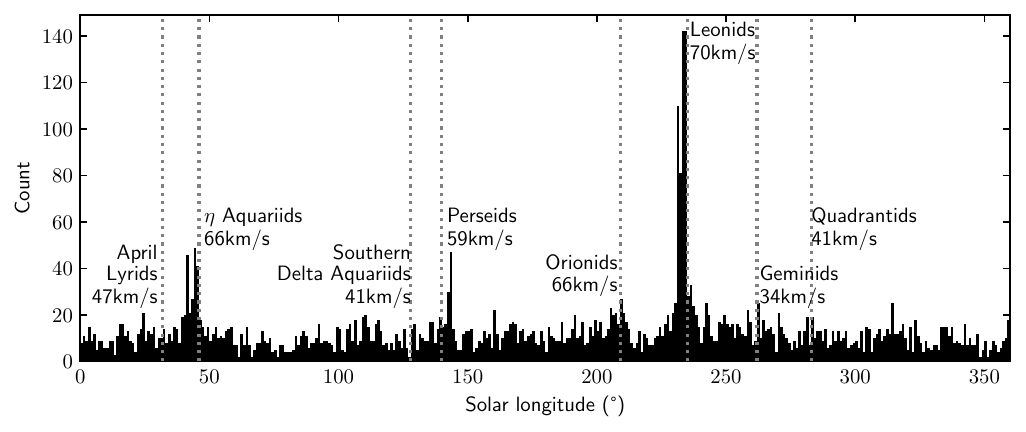}
        \caption{GLM bolide detections in the \texttt{machine} database (2019-07-01 to 2022-06-03) by the solar longitude at the time of the impact, with $0^\circ$ being the vernal equinox. Dotted lines represent the solar longitudes of the peaks of different meteor showers \citep{showers}.}
        \label{fig:sollon}
\end{figure*}

The final detail to work out is that we are interested in this latitudinal variation as a window into the validity of models (e.g.~that of \cite{darrel}) which predict it for large impactors of planetary defense interest. As all detections made so far by GLM are of small and non-hazardous objects, it is a leap to directly compare both populations. In particular, their orbital distributions may differ due to the different dynamical effects acting at different size ranges. For instance, the Yarkovsky effect may be stronger for the large meteoroids and large asteroids that GLM detects, though it also depends on other physical characteristics \citep{bottke2006}. Nevertheless, we can assume for our discussion that the orbital distribution of large sporadic meteoroids and smaller asteroids ejected onto Earth-impacting trajectories is similar to that of larger, hazardous impactors. The same cannot be said for meteor showers, which arise from large groups of objects with similar, cometary orbits and higher radiant velocities than asteroidal NEOs.

As seen in Figure \ref{fig:sollon}, a histogram of detections by solar longitude, some meteor showers are quite prominent in GLM data. This has proven useful for studies of meteor showers and their properties such as \cite{kasia}, but in this case the presence of meteor showers makes it impossible to treat GLM data as solely a reflection of the population of sporadic meteoroids and small asteroids. For this reason we design ways in Section \ref{sec:methods} to account for the presence of meteor showers when estimating the latitudinal variation. We do this without assigning individual detections to meteor showers as GLM data lacks the information---precise radiant and velocity---typically used to do so.

\subsection{How to theoretically model the latitudinal variation}
\label{sec:theoretical-flux}

\begin{figure*}
    \centering
    \includegraphics[width=0.49\textwidth]{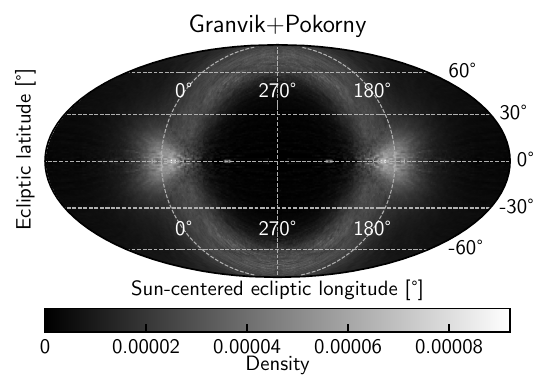}\hfill
    \includegraphics[width=0.49\textwidth]{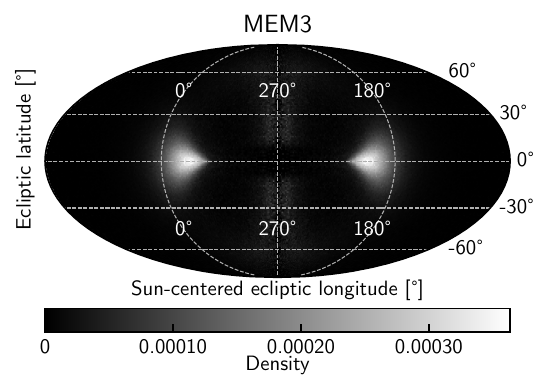}\\\vspace{-0.7em}
    \includegraphics[width=\textwidth]{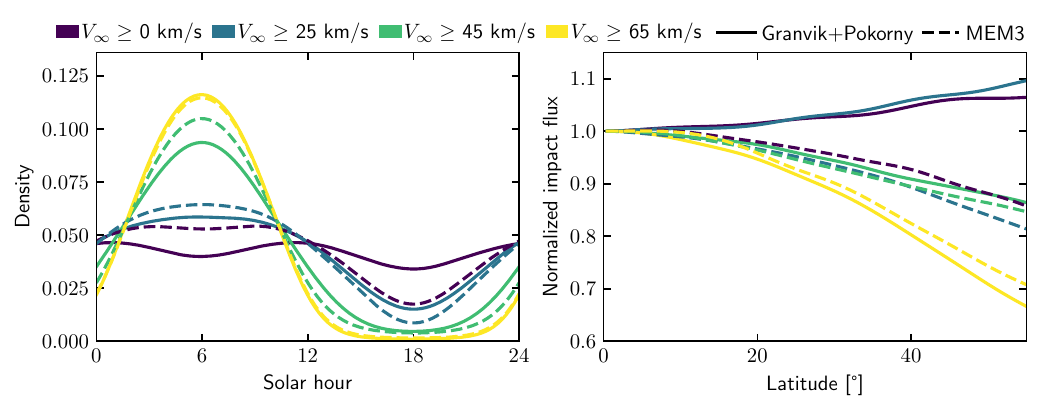}
    \caption{Top: Joint density of impact radiant ecliptic latitude and Sun-centered ecliptic longitude computed using (left) the Granvik+Pokorný method described in Section \ref{sec:theoretical-flux} and (right) the Meteoroid Engineering Model 3 with a circular orbit at \qty{1}{AU} \citep{mem}. In this coordinate system, the Sun is at (0, 0) and the direction of the Earth's motion---the Apex---is at (270, 0). The Helion and Antihelion sources are clearly visible at roughly (0,0) and (180,0), respectively. In the Granvik+Pokorný plot, the small clusters in the ecliptic plane between these two sources may be due to clusters of objects on similar orbits in \cite{granvik} or due to the numerical methods used by \cite{pokorny}. Bottom: Theoretical impact flux by (left) the local solar hour at the time of impact and (right) the latitude, computed using the extension of the methods of \cite{darrel} described in Section \ref{sec:theoretical-flux}. The values in the latitude plot are per unit area. A solar hour of 12 corresponds to noon in solar time, and a solar hour of 6 corresponds to, for an observer at the equator, when the direction of the Earth's motion is directly above (at zenith). Even though they model very different size ranges of objects and have different distributions of radiants, Granvik+Pokorný and MEM3 produce similar results when impact locations and times are simulated from their radiant distributions.}\label{fig:eclon-eclat}
\end{figure*}

\noindent To contextualize and understand empirical results from GLM, we must leverage previous works which model impact distributions. We follow the work of \cite{darrel} with some minor changes. Broadly, the pipeline for computing a theoretical impact distribution across latitudes (or, for that matter, any other geographic statistic of impacts), relies on three steps:

\begin{enumerate}
    \item Model the orbital distribution of objects which might impact the Earth. The output of this step can be a large sample of synthetic NEO orbits. We follow \cite{darrel} in using the results of \cite{granvik} in this work. Specifically, we use their published debiased set of 802,000 synthetic NEO orbits for objects of absolute magnitude $17\leq H\leq 25$.
    \item Compute impact probabilities for each of the several ways objects on such orbits may impact the Earth. This is done in works such as \cite{pokorny}, whose code for computing impact probabilities and radiants we use in this work, following \cite{darrel}. For our purposes, the output of this step is a three-dimensional joint probability distribution across impact radiant velocity, ecliptic latitude, and Sun-centered ecliptic longitude. Figure \ref{fig:eclon-eclat} displays this probability density marginalized over velocity, showing the Helion and Antihelion sources clearly. Including Sun-centered ecliptic longitude is an addition over \cite{darrel}.
    \item Using a sample of impact radiant velocities, ecliptic latitudes, and Sun-centered ecliptic longitudes, compute the geographic distribution of impacts. \cite{darrel} discuss how to obtain the variation across latitudes using a Monte Carlo simulation. We extend the method slightly to estimate diurnal variation, which is enabled by our inclusion of Sun-centered ecliptic longitude. Figure \ref{fig:eclon-eclat} displays some results from this step.
\end{enumerate}

\noindent To explain our extension in the third step, we must first briefly describe the Monte Carlo simulation of \cite{darrel}. For a given radiant velocity, the ``impact disk'' is a region of space such that any objects passing through it will impact the Earth, and any not passing through it will not. For each sampled radiant velocity and ecliptic latitude, the simulation samples a uniformly random point on the impact disk. Then, the simulation analytically computes an impact point in a generic coordinate system, and rotates it according to ecliptic latitude. The simulation then samples the angle between the direction of the Earth's axial tilt and the impactor as well as a uniformly random hour angle. Finally, through a series of coordinate transformations, the geographic latitude and longitude of the impact are computed.

This method, by averaging over hour angles, cannot yield the diurnal variation which we would like to estimate. Hence we modify it slightly. Specifically, we follow it up to the point of obtaining an impact point in a generic coordinate system. But instead of randomly choosing an hour angle and an angle relative to the Earth's tilt, we use a randomly chosen date and time along with the sample's Sun-centered ecliptic longitude to transform the coordinates and obtain geographic coordinates. The randomly chosen date and time, when combined with the computed longitude, allows local solar time to be computed. We think this slight addition to the method of \cite{darrel} is promising as it incidentally also allows many other interesting variables to be computed, like the solar altitude at the time and location of impact. In principle various seasonal variations could be computed, as well as the theoretical diurnal variation at a specific ground-based observing site on a specific date. In this work we only pursue this method so far as it informs our analysis of GLM data, though it could be taken much further.

Following Steps 1-3 end-to-end by plugging the results of one step into the next yields a theoretical geographic distribution of impacts from all sources modeled in the first step, and this is done by \cite{darrel}. But in addition to the modifications which allow solar hour to be computed, we can simply adjust the sample taken in Step 3 to obtain other results of interest.

One such modification is selecting a subset of radiant velocities when sampling (effectively conditioning the samples). As will become apparent, to properly contextualize GLM data we need to know how the resulting geographic distributions change if we only sample impacts with high radiant velocities.

GLM is sensitive to meteor showers, as discussed in Section \ref{sec:how} and illustrated in Figure \ref{fig:sollon}. Hence the theoretical latitudinal variation of individual meteor showers is also of interest. To compute it, we can simply fix the radiant velocity, ecliptic latitude, Sun-centered ecliptic longitude, and the sampled date in Step 3 to the appropriate values for a given meteor shower. We obtain meteor shower parameters from \cite{showers}. Since the assumption of North-South impact symmetry no longer applies for specific meteor showers as it does for the total NEO population, we do not fold the distribution to symmetrize it as is done in \cite{darrel}.

The final modification we make to this method is that, after completing Step 3, we can select a subset of ``valid'' samples. In the context of GLM we can, for instance, only keep samples that fall within the GLM field of view. This allows for a more direct comparison between modeled distributions and simple histograms of GLM data, but of course it does not account for GLM's other detection biases.

As discussed in Section \ref{sec:how}, the smaller objects detected by GLM might have a different distribution of orbits than the larger objects modeled in works like \cite{granvik}. The method of \cite{granvik} does allow for extrapolation to smaller objects with a higher absolute magnitude of $H>25$, yet it only does so with the assumption that they have the same orbital distribution as objects with $H=25$. So we rely on their database of 802,000 synthetic orbits. The more recent NEOMOD model \citep{neomod} models the orbital distribution of objects as dim as $H=28$, which would get closer to the population that GLM detects, but it does not include the argument of perihelion as a parameter and thus cannot be directly used with the methods of \cite{pokorny} without further modeling.

Nonetheless, we would like to understand how our choice of models in Steps 1-2 (which we can collectively call the Granvik+Pokorný model) affects how we interpret GLM data. For comparison to a model with far smaller particle sizes, we also use the Meteoroid Engineering Model 3 (MEM3) from \cite{mem} to obtain the joint density of radiant velocity, ecliptic latitude, and Sun-centered ecliptic longitude. Results from it are shown alongside those from the Granvik+Pokorný model in Figure \ref{fig:eclon-eclat}. The two models have very different radiant distributions, with MEM3 having far stronger Apex and Toroidal sources (almost nonexistent in Granvik+Pokorný). As seen in Figure \ref{fig:eclon-eclat}, this difference produces quite different diurnal and latitudinal variations when all radiant velocities are taken into account. But when restricting impactors to those with high radiant velocities, the results are quite similar, and we are confident that our choice of model does not affect how we interpret GLM data.

\subsection{Indications of a velocity bias}
\label{sec:vel_ind}

\noindent Previous works have examined the increase in emissions in the \qty{777}{\nm} band as a function of velocity \citep{vojacek}, and it would stand to reason that, all else being equal, the faster a bolide is the easier it is for GLM to detect it. In fact, this effect appears to be exaggerated in GLM compared to traditional meteor cameras. \cite{vojacek} argue in their analysis that the limiting visual magnitude for GLM is $-14$ for bolides at \qty{15}{\km\per\s}, but $\approx -8.8$ for bolides at \qty{70}{\km\per\s}. This means that bolides which have a relatively dim visual magnitude but are fast enough can still be detected by GLM. Estimating the velocities of bolides detected by GLM is nontrivial as most are only detected by one satellite. For these bolides, even if it is possible to determine their trajectories through a plane parallel to that of the sensor, it is not possible to determine their movement towards or away from the satellite, so a velocity cannot be easily obtained. Even for bolides detected by two satellites, the large pixel pitch (at minimum, equivalent to a distance of \qty{8}{km} on the Earth's surface) means that most bolides are only detected in a few pixels. Estimating velocities with uncertainties using GLM data is the subject of future work. Nevertheless, without knowing the velocities of the detections, there are still indications in the data itself that GLM has a strong detection bias towards high-velocity objects.

Looking closer at Figure \ref{fig:sollon}, GLM especially sees increases in flux around the peaks of the Leonids, Perseids, and $\eta$ Aquariids. There also appears to be an increase at the Orionids peak. Though it is not yet possible to attribute GLM detections to specific meteor showers, this temporal correspondence suggests that GLM is particularly sensitive to these showers. Notably, these are all showers with relatively high radiant velocities: from \qty{59}{\km\per\s} for the Perseids to \qty{70}{\km\per\s} for the Leonids \citep{showers}. Other showers which are prominent in ground-based data, like the Quadrantids, Geminids, April Lyrids, and Southern Delta Aquariids, do not seem to be associated with an increase in flux in GLM data. These all have far lower radiant velocities (from \qty{34}{\km\per\s} for the Geminids to \qty{47}{\km\per\s} for the April Lyrids) than the showers which GLM appears most sensitive to.

Of course, these meteor showers may have differences in their flux of larger meteoroids. The Leonids shower in particular is known to have especially large particles \citep{trigorodriguez}. Another consideration is that the flux of larger meteoroids in a meteor shower can have a far narrower peak than the flux of smaller meteoroids in the same shower \citep{jenniskens2006}. As GLM's field of view is not global, it is therefore possible for both GLM sensors to be on the wrong side of the Earth to observe the peak in larger meteoroid impacts in a given year. This could combine with the general year-to-year variability in flux from meteor showers and lead to a large yearly variability in meteor shower flux observed by GLM. This has already been noted for the Leonids, which GLM appears to have detected far more of in 2020 than in 2019 \citep{clark}. Only a few years of GLM bolide detections are available, and hence the data presented in Figure \ref{fig:sollon} might not be a representative sample of GLM's ability to detect different meteor showers. Nevertheless, the relative prominence of meteor showers with higher radiant velocities is highly suggestive of a velocity bias.

\begin{figure*}
    \includegraphics[width=\textwidth]{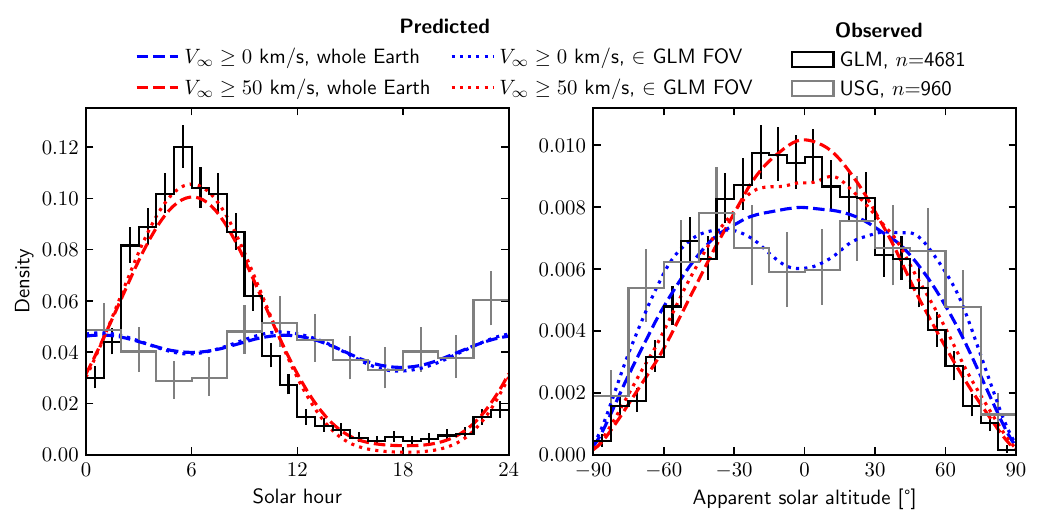}
    \caption{Bolide detections by (left) the local solar hour at the time of impact and (right) the apparent solar altitude. A solar hour of 12 corresponds to noon in solar time, and a solar hour of 6 corresponds to, for an observer at the equator, when the direction of the Earth's motion is directly above (at zenith). Apparent solar altitude is the altitude, in degrees, of the Sun above the horizon assuming no atmospheric refraction. Density data are obtained from the \texttt{machine} database for GLM and, for USG, all published USG detections as of 2023-07-26 are used. In the left plot, it is apparent that GLM sees a strong diurnal variation which peaks around 06:00, while USG sees a weaker diurnal variation with peaks at 00:00 and 12:00. The theoretical curves for different radiant velocities $V_\infty$ are obtained using an extension of the method of \cite{darrel}, as described in Section \ref{sec:theoretical-flux}. The errors on the USG and GLM densities are $1\sigma$ confidence intervals on the rates of the Poisson distributions generating the counts (these are not exactly $\sqrt{N}$ Poisson errors).}
    \label{fig:solar}
\end{figure*}
\begin{figure}
    \centering
    \includegraphics[width=0.5\textwidth]{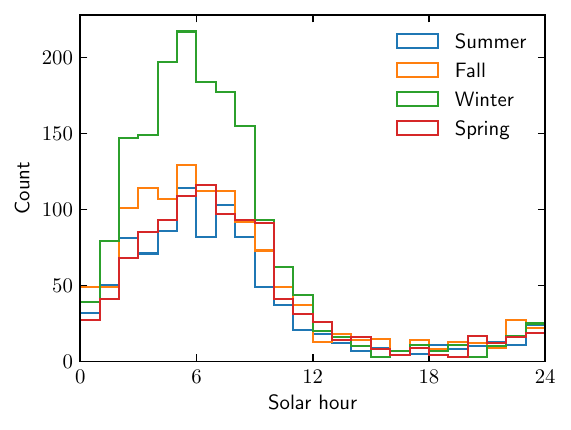}
    \caption{GLM bolide detections in the \texttt{machine} database by the local solar hour at the time of the impact (12 being noon), split apart by season. A solar hour of 6 corresponds to when, for an observer at the equator, the direction of the Earth's motion is directly above (at zenith). Detections are assigned to ``seasons'' according to the solar longitude at the time of the detection (with $0^\circ$ being the vernal equinox). Spring corresponds to a solar longitude in the interval $[-45,45]$, Summer to $[45,135]$, Fall to $[135, 225]$ and Winter to $[225,315]$. These differ from traditional calendar seasons. The extreme diurnal variation is present in all seasons, but is especially pronounced in the winter, when the Leonids meteor shower occurs.}
    \label{fig:solar-season}
\end{figure}

The other clue that GLM is much more sensitive towards faster objects is the diurnal variation of the detections. Figure \ref{fig:solar} shows that, in local solar time at a bolide's location, detections peak at six in the morning in the GLM data. This is the case in all seasons, as seen in Figure \ref{fig:solar-season}, which indicates that this diurnal variation is not solely due to meteor showers like the Leonids and Perseids. Six in the morning is also the local solar time when a standing observer at the equator is travelling head-first in the direction of the Earth's motion. USG data does not show such a peak. Given that, in both data sets, there does not appear to be a major effect of solar illumination on detection rates, this again suggests a detection bias towards faster objects in GLM data, as, generally, the fastest objects (in terms of relative velocity to the Earth) come from retrograde orbits and hence will tend to impact at around 06:00.

In fact, Figure \ref{fig:solar} also includes diurnal variation estimates obtained using the method described in Section \ref{sec:theoretical-flux}. It demonstrates that the diurnal variation observed by GLM is indeed far more consistent with predictions for modeled impactors having a minimum radiant velocity of \qty{50}{\km\per\s} than for the total population of modeled impactors. The estimates also clearly demonstrate that, if there were no velocity bias, we would expect a bimodal distribution with peaks around noon and midnight in local solar time. These peaks correspond to radiants in the Helion and Antihelion sources, respectively, and such a bimodal distribution is seen in the USG data. The detection threshold in the USG data is much higher (almost every USG detection in GLM's FOV is detected by GLM, but most GLM detections are not detected by USG), which may effectively filter out smaller objects such as those in meteoroid streams. Different sensor characteristics, such as USG's use of broadband sensors \citep{tagliaferri} as opposed to GLM's narrowband sensors, may also affect how sensitive USG sensors are to bolide velocity.

GLM's distribution over solar hour can also be compared to the diurnal variation of meteor rates as detected by ground-based meteor radars (which, unlike cameras, can make detections during the day). The ratio of GLM detection counts from the peak at 05:30-06:30 to the trough at 17:30 to 18:30 is $516/32=16.125$. This is very much on the extreme end of the diurnal variations reported by several meteor radar systems in various locations and times of year \citep{lovell, okamoto, singer, szasz}, which again suggests that GLM has a strong bias towards high-velocity objects.

We must be careful about taking these diurnal variation measurements beyond their use here as an indication that GLM has a bias towards high-velocity objects (which themselves generally come from the Apex source). Again, according to \cite{vojacek}, GLM's limiting magnitude for bolides varies widely for different velocities, and these objects coming from the Apex source could be smaller and dimmer in visual magnitude than the decimeter- and meter-sized objects GLM was initially thought to detect.

\section{Methods}\label{sec:methods}

\subsection{Poisson Regression Model}\label{sec:poisson}
\begin{figure*}
        \begin{center}
        \includegraphics[width=\textwidth,page=1]{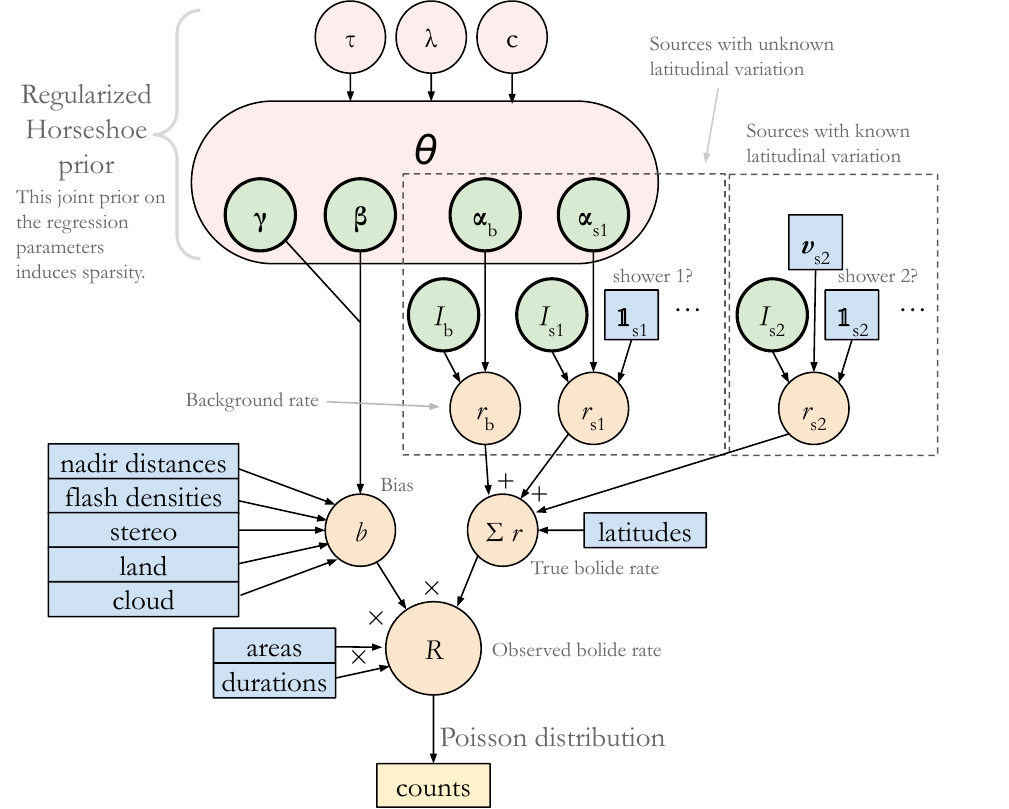}\end{center}
        \caption{Diagram of the Bayesian Poisson regression model developed in Section \ref{sec:poisson}. Shapes with rounded corners are free parameters, and rectangles are data obtained from the discretization described in Section \ref{sec:disc}. An arrow $A\to B$ indicates that $B$ conditionally depends on $A$. Of course, if $B$ conditionally depends on $A$, then $A$ conditionally depends on $B$ (and there are many more implied conditional dependencies in this diagram), but the arrows show how the variables are directly defined as depending on each other. Bold outlines indicate the coefficients we wish to estimate. The regularized horseshoe prior, discussed in Section \ref{sec:bayesian}, is a joint prior on all of the coefficients except for intercepts $I$. Counts are generated according to the Poisson distribution.}
        \label{fig:model}
\end{figure*}
\noindent We are investigating bolide flux as a function of latitude, with angle of incidence and other biases accounted for. We can assume that impacts are independent in the sense that one impact happening in a location at some time does not raise or lower the probability of another impact happening in another location at another time. We can assume this even while treating the flux of bolides as different across space and time. This independence would be violated if, say, a sporadic meteor broke up just before reaching the Earth and resulted in several impacts clustered in space and time, but there is no indication that something like this is common in GLM data. Because of this independence assumption, we can model the observed bolides as coming from a Poisson distribution with an inhomogeneous rate depending on latitude and some biases. The model can then be expressed as:
\begin{equation}N(L,T)\sim \pois\left(\int_T\int_LR(l,t)\ dl\ dt\right), \label{eq:pois}\end{equation}
\noindent where $N(L,T)$ is the number of bolides in an area $L$ (a part of the Earth's surface) over a set of times $T$, and $R(l, t)$ is the inhomogeneous rate function (in terms of $l$, location, and $t$, time) of the Poisson distribution that generates $N(L,T)$. $l$, the location, can be thought of as a $(\phi,\lambda)$ longitude-latitude tuple or some other representation of location on the Earth's surface. $N(L,T)$ lies at the bottom of Figure \ref{fig:model} as the observed counts. Throughout this section, Figure \ref{fig:model} can act as a guide to the relationships between the variables.

The rate of bolides observed by GLM is the sum of location-dependent rates of bolides coming from different sources, multiplied by a location-dependent bias term which reduces detection efficiency. So we can represent $R$ thus: \begin{equation}R(l,t) = b(l) \sum_{s\in S(t)}{r_s(l)},\label{eq:R}\end{equation}
where $b(l)$ is the bias term and $S(t)$ is a set of bolide sources active during time $t$. The bias term $b(l)$ does not depend on time as variation in rates due to showers is captured in the sum of rates, and we ignore the hour-scale diurnal variation in detections discussed in Section \ref{sec:vel_ind}. We define $S(t)$ as always including a source representing the background rate of bolides. Depending on the time $t$, $S(t)$ may also include meteor showers as a source. While meteor showers and the background do vary in rate over time, this model approximates the rate as constant in time while a source is active. The intention is not to study the variation in time, but to approximate away the elevated rates (with different latitudinal distributions) associated with the presence of a meteor shower. For a bolide source $s$, $r_s(l)$ represents the rate attributed to a particular bolide source as a function of location.

$b(l)$ and $r_s(l)$ are both modeled as exponentiated polynomials. The exponentiation is necessary to ensure that $R(l,t)$ is positive, as a rate cannot be negative. Let $\angle(l)$ be the angle of incidence of light from location $l$ into the sensor of the satellite in question (see Figure \ref{fig:aoi}). $\angle(l)$ can be computed by finding the 3-dimensional position vectors $\B p$ of the satellite and $\B q$ of $l$, obtaining the ``look vector'' from the satellite to the point on Earth as $ -\B p + \B q$, and finally taking the angle between this look vector and $\B p$ (or, more intuitively, $\B -\B p$). $\B p$ and $\B q$ can be obtained in any reasonable Earth-centered reference frame. As discussed in Section \ref{sec:how}, the angle of incidence affects the amount of light received through the filters, and hence we expect it to affect bolide detection efficiency.

Let $\mathrm{Flash}(l)$ be the lightning flash density at location $l$, expressed in units of \unit{km^{-2}.year^{-1}}. This data is obtained from two previous lightning imaging systems: the Lightning Imagining Sensor and the Optical Transient Detector \citep{flashdensity}. Let $\mathrm{Cloud}(l)$ be the cloud fraction at location $l$ averaged over one year of Terra MODIS data, with 1 meaning $l$ was constantly cloudy and 0 meaning $l$ experienced no cloud cover \citep{modis}. Finally, let $\mathrm{Land}(l)$ be 1 if location $l$ is over land, and 0 otherwise, and $\mathrm{Stereo}(l)$ be likewise for the GLM stereo region (the intersection in Figure \ref{fig:detections}). We define the bias term as:

\begin{equation}b(l) = \exp\left(\gamma_{\text{flash}} \mathrm{Flash}(l) + \gamma_{\text{cloud}}\mathrm{Cloud}(l) + \gamma_{\text{land}} \mathrm{Land}(l) + \gamma_{\text{stereo}} \mathrm{Stereo}(l) + \sum_{i=1}^{k_d} \beta_i(\angle(l))^i\right)\label{eq:b}\end{equation}
For the $\texttt{machine}$ database, we fix $\gamma_{\text{stereo}}=0$ as, unlike in $\texttt{human}$, there ought not be any major bias towards the stereo region in $\texttt{machine}$.

Let the latitude of a location $l$ be $\lambda(l)$. In the long-run, latitude should be the only component of location affecting the true rate of bolide impacts, as the rotation of the Earth averages the rate over longitudes. Hence the rate function ought to only depend on $l$ through $\lambda(l)$. There are two possibilities for the rate function $r_s(l)$: we know the latitudinal variation of $s$, or we don't. If it is unknown, we define $r_s(l)$ as the following polynomial:
\begin{equation}r_s(l) = \exp\left(I_s+\sum_{i=1}^{k_{\lambda}}\alpha_{s,i}(\lambda(l))^i\right).\label{eq:r}\end{equation}
Unlike the polynomial in the bias term, there is an intercept $I_s$ here. $k_d$ and $k_\lambda$ are numbers which define the degree of the polynomials used. We set $k_d=k_\lambda=6$ to allow for a wide range of possible functions.

If we do know the latitudinal variation of $s$---as we do for meteor showers---we still do not know the actual rate with which GLM detects that shower. Let $v_s(\lambda(l))$ be a function representing the flux (relative to the equator) of source $s$ at a location $l$ on the Earth, which in practice will only depend on $l$ through its latitude $\lambda(l)$. Then:
\begin{equation}r_s(l)=I_sv_s(\lambda(l)).\label{eq:r-known}\end{equation}
Here, $I_s$ acts as a free parameter to scale the known latitudinal variation.

Expressing the model in this way allows for the simultaneous estimation of the detection bias as a function of angle of incidence and the additive rates due to different meteor showers and the background rate. As specified in Equation \ref{eq:R}, multiplying the detection bias by the sum of the rates gives the total observed rate of GLM. Note that, while all of the detection biases are based on geographic location, none of them are solely dependent on latitude. Angle of incidence, for instance, is rotationally symmetric, and points on the extreme edges of a satellite's field of view in latitude or in longitude will have a high angle of incidence.

\subsection{Discretization}\label{sec:disc}
\begin{figure*}
        \includegraphics[width=\textwidth]{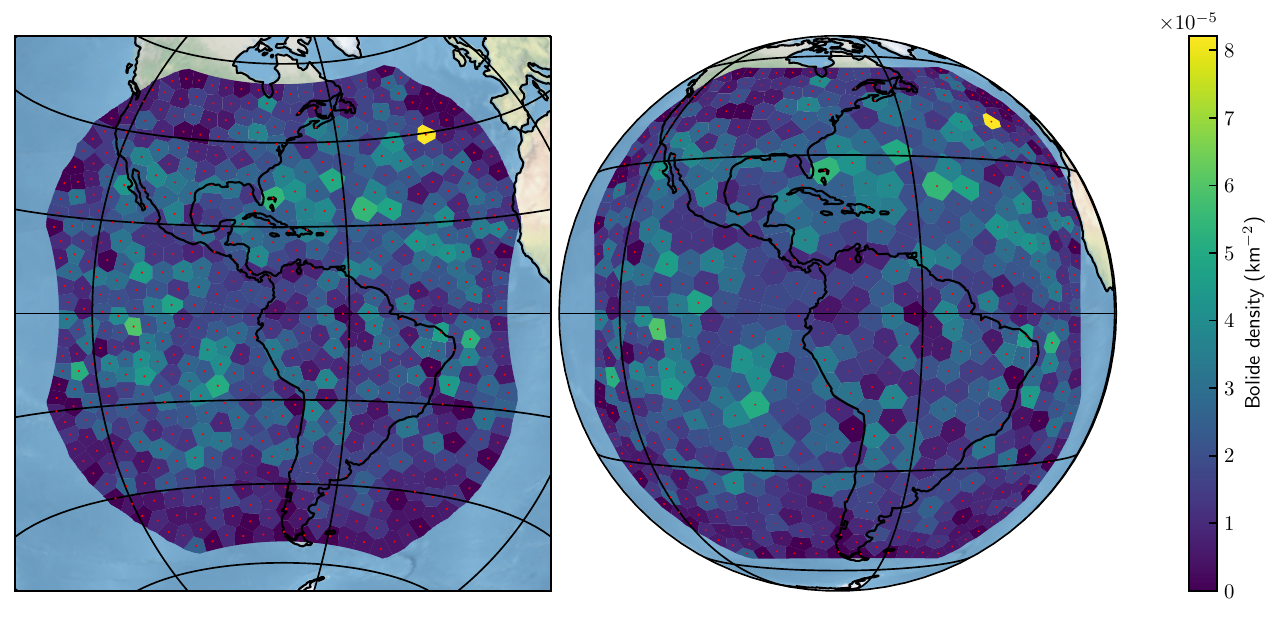}
        \caption{Sample discretization of GOES-16's bolide detections. The field of view is split into small polygons, and the density of bolides in each polygon is shown. The red points are the centroids of the polygons, which are used to compute the latitude, longitude, and angle of incidence. Left: The Azimuthal Equal Area projection centered on GOES-16's nadir, which is the space where the discretization is computed. The polygons are roughly uniform in shape and size, including at the edge of the FOV. Right: Globe from the perspective of GOES-16. The GLM field of view, though it takes on an odd shape in most projections, is really a rectangle with the corners masked. The bright polygon is an outlier due to randomness (both in the observed locations of the bolides and in the random discretization).}
        \label{fig:discretize-density}
\end{figure*}

\begin{algorithm}\caption{Discretization}\label{alg:discretization}
\begin{algorithmic}[1]
\STATE Let $P$ be a uniformly random set of points within the satellite's FOV.
\FOR{$i\in 1,\ldots,k$}
    \STATE Compute the Voronoi diagram $V$ from $P$, \citep{scipy, qhull} represented as a set of polygons.
    \STATE Clip each polygon of $V$ to the sensor's FOV.
    \STATE $P\gets$ the centroids of the polygons of $V$
\ENDFOR
    \STATE Count the number of detections in each polygon of $V$.
    \STATE Compute the latitude, longitude, and angle of incidence of each polygon in $V$ according to its centroid.
    \STATE Compute the area of each polygon.
    \STATE Compute the lightning flash density of each polygon, its average cloud fraction, whether or not it is above land, and whether or not it is in the stereo region.
    \STATE Compute the relative observation duration of each polygon.
\end{algorithmic}
\end{algorithm}
\noindent The spatial Poisson regression model we use has a computationally intractable likelihood function, as in principle the presence or absence of a detection in each infinitesimal region must be considered. Hence we must discretize the point data into a set of counts for a finite number of regions. It is possible to simply grid the data, but we instead use a randomized method which allows us to sample from the model multiple times, with different random polygons, and be confident that an arbitrary gridding is not affecting the results. For each satellite (GOES-16 and GOES-17), we apply Algorithm \ref{alg:discretization}. In each step, the Azimuthal Equal Area projection centered at the satellite's nadir is used as the space in which geometry occurs. An equal area projection means that there is no systematic effect on polygon size of latitude, longitude, or angle of incidence. The rotational symmetry of the Azimuthal Equal Area projection in particular ensures that there are no differences between the general shapes of the polygons made at the extreme edge of the FOV in longitude and at the extreme edge in latitude (see Figure \ref{fig:discretize-density}).

The loop in Step 2 of the algorithm approaches a centroidal Voronoi diagram, which creates polygons that are uniform in size within a boundary \citep{voronoi}. In principle, it is not actually necessary to use polygons of uniform size, but doing so guarantees a fine resolution across the field of view with fewer polygons. Clipping the polygons to the FOV ensures that the shape of the FOV is taken into account so that, for instance, the lack of observations in the extreme North and South just outside the FOV isn't taken to imply a decrease in flux at high latitudes. We emphasize that, when iterated, this process of clipping still yields polygons that are uniform in size.

The ``relative observation duration'' is 1 for most polygons. But GOES-17 performs semiannual yaw flips, slightly changing its FOV \citep{yawflip}. For those polygons which are only observed for half of the year by GOES-17, the relative observation duration is set to $1-a/2$, where $a$ is the proportion of the polygon which is only in GOES-17's inverted FOV or non-inverted FOV (i.e.~the proportion of the polygon within the symmetric difference of the inverted and non-inverted FOV). This ensures that fewer detections in the regions where GOES-17 only observes for half of the year, which also happen to be at extreme latitudes, are not taken to imply a decrease in flux at high latitudes.

This discretization works because, within a small, convex polygon, location will (naturally) vary little, and $R$ will be roughly constant. Thus we get the relation that:
\[\int_L R(l,t)\ dl\approx \text{Area}(L) R(\text{Centroid}(L), t)\]
Where $\text{Area}(L)$ is the area of polygon $L$ and $\text{Centroid}(L)$ is the location of the polygon's centroid.

For additionally discretizing time, since we assume that bolide sources are either active or not active given a time, we see that, if $S(t)=S(t_0)\ \forall t\in T$, then:
\[\int_T \text{Area}(L) R(\text{Centroid}(L),t)dt=\mu(T) \text{Area}(L)R(\text{Centroid}(L),t_0),\]
where $\mu(T)=\int_Tdt$, i.e.~the total duration of $T$. In other words, if, for any time $t$ in $T$, the bolide sources active are the same as those active at time $t_0$, then we only need to consider $R$ at $t_0$. So, in practice, Equation \ref{eq:pois} can be approximated as:
\begin{equation}N(L,T)\sim \pois\left(\mu(T)\text{Area}(L)R(\text{Centroid}(L),t_0)\right), \label{eq:pois-approx}\end{equation}
if $L$ is small (true with enough polygons) and $S(t)=S(t_0)\ \forall t\in T$ (true if only considering the background source).

We do not always wish to only consider the background source. So, within each polygon produced by the discretization, we group the data into bolides within each target meteor shower, and bolides not associated with any of the target meteor showers. If there are $k$ total meteor showers modeled, we are effectively left with $k+1$ final polygons for each original polygon (assuming no meteor showers intersect in time). Each of the $k+1$ polygons has the same shape as its original polygon, but only counts the bolides corresponding to its bolide source. We add $k$ new indicator variables to the polygons indicating which showers they belong to.

We are thus left with a table of polygons with several features filled in for each polygon. Namely: its count of bolides; its centroid's latitude, angle of incidence, and area; its relative observation duration; its lightning flash density, cloud coverage, whether or not it is on land, and whether or not it is in the stereo region; and finally, several indicator variables representing which meteor showers it belongs to. All of these variables are represented by rectangular shapes in Figure \ref{fig:model}. Importantly, each of the polygons created satisfies the conditions for Equation \ref{eq:pois-approx} to hold. The likelihood of the count of each polygon is now trivial to compute as it has a single rate, and not a rate function. The likelihood of the whole dataset under a parameter setting can then be computed by taking the product of the likelihoods of the counts in each polygon.

\subsection{Bayesian representation and regularization}\label{sec:bayesian}
\noindent The above model could be fit by simply selecting the $\alpha$'s, $\beta$'s, $\gamma$'s, and $I$'s in Equations \ref{eq:b} and \ref{eq:r} to maximize the likelihood of the observed data. However, this would not allow for proper uncertainty quantification. The model is rather complicated, and we would especially like to state uncertainties on the joint fit of the $\alpha$'s (the parameters of the polynomial on latitude). We are more interested in the values taken by the polynomial at different latitudes than in the individual parameters.

To accomplish this, we take a Bayesian approach. We give any intercepts $I_s$ lying within an instance of Equation \ref{eq:r} a Cauchy(0.5) prior, and those lying in an instance of Equation \ref{eq:r-known} a HalfCauchy(0.5) prior (to force positivity). Using the Cauchy and HalfCauchy distributions as priors is appropriate here as they have heavy tails, encoding a lack of prior knowledge on these scaling factors. We use the regularized-horseshoe prior \citep{reg-horseshoe} as a joint prior on the $\alpha$'s, $\beta$'s, and $\gamma$'s, with the data pertaining to each coefficient normalized to have a mean of zero and a variance of one. This prior is illustrated at the top of Figure \ref{fig:model}.

More precisely, for $\tau$, the ``global shrinkage parameter,'' we say:
\[\tau \sim \mathrm{HalfCauchy(1)}.\]
For $\boldsymbol\lambda$, the ``local shrinkage parameters,'' we say:
\[\lambda_i\sim \mathrm{HalfStudent\text{-}t}_{2}.\]
For $c^2$, the ``slab width,'' we say:
\[c^2 \sim \mathrm{Inv\text{-}Gamma}(1,1).\]
Finally, if we say:
\[\boldsymbol{\tilde{\lambda}}=\frac{c^2\boldsymbol\lambda}{c^2+\tau^2\boldsymbol\lambda^2},\]
then we can let our parameter vector $\boldsymbol\theta$ which represents the $\alpha$'s, $\beta$'s, and $\gamma$'s together be distributed according to:
\[\theta_i \sim \mathrm{N}(0,\tau^2\tilde{\lambda}_i^2).\]
We thus largely follow the recommendations of \cite{reg-horseshoe}, except for our highly uninformative prior on $\tau$. We use this prior as, given how the model is formulated, it is not exactly a generalized linear model when meteor showers are included, and hence there is currently no plug-in result for a good informative prior on $\tau$ given a guess for the true number of parameters. We find that there is enough data in this study to constrain $\tau$ tightly even with the uninformative prior.

Using this regularized horseshoe prior pushes the model towards sparsity, which allows us to include all possibly relevant parameters in the Poisson regression without giving too high a prior probability to extreme models where each coefficient is large. This helps tighten the resulting posterior distribution and improves sampling. We sample this posterior distribution using Markov Chain Monte-Carlo (MCMC), which is a method to do a random walk across possible assignments of variables in a way that, in the long-run, the samples approximate the posterior distribution of those variables. Specifically, we use the No-U-Turn Sampler \citep{nuts} as implemented in PyMC \citep{pymc}. We use 4 chains with 5,000 burn-in samples and 10,000 samples each, and verify that each run converged to the posterior using standard metrics like $\hat{R}$ and Effective Sample Size \citep{rhat}.

\section{Results}\label{sec:results}

\begin{table*}
    \caption{The models tested. ``Threshold'' refers to the confidence threshold used when selecting data from the GLM bolide-detection pipeline. ``Showers'' refers to the meteor showers included as bolide sources---the inclusion of a shower as a bolide source $s$ means that for a time $t$ 5 days around the shower peak, $s\in S(t)$ in Equation \ref{eq:R}. ``Shower variation known'' reflects whether the latitudinal variation of the showers is treated as unknown and is estimated via Equation \ref{eq:r}, or is hard-coded based on theoretical estimates, at which point Equation \ref{eq:r-known} is used. Model 1 serves as the reference model which other models are compared to.}\label{tab:models}
\centering
\begin{tabular}{c|cccccc}
\textnumero & Data & Threshold & Showers & Shower variation known & Stereo bias & Satellite(s)\\\hline
1 & \texttt{machine} & 0.7 & - & - & N & G16, G17\\
2 & \texttt{human} & - & - & - & Y & G16, G17\\
3 & \texttt{machine} & 0.7 & - & - & N & G16 \quad\quad\ \\
4 & \texttt{machine} & 0.7 & - & - & N & \quad\quad\  G17\\
5 & \texttt{machine} & 0.7 & LEO, PER & N & N & G16, G17\\
6 & \texttt{machine} & 0.7 & LEO, PER & Y & N & G16, G17\\
7 & \texttt{machine} & 0.25 & - & - & N & G16, G17\\
8 & \texttt{machine} & 0.5 & - & - & N & G16, G17\\
9 & \texttt{machine} & 0.8 & - & - & N & G16, G17\\
10 & \texttt{machine} & 0.9 & - & - & N & G16, G17\\
11 & \texttt{machine} & 0.95 & - & - & N & G16, G17\\
12 & \texttt{machine} & 0.99 & - & - & N & G16, G17\\
\end{tabular}
\end{table*}
\begin{table}
\caption{The figures shown, which models they correspond to, and which estimated parameters are displayed in each.}\label{tab:figures}
\centering
\begin{tabular}{r|rl}
Figure \textnumero & Models & Parameters\\\hline
\ref{fig:machine-lat} & 1 & $r_{\text{base}}(l)$ \\
\ref{fig:human-lat} & 2 & $r_{\text{base}}(l)$, $\gamma_{\text{stereo}}$\\
\ref{fig:fov} & 1, 2 &  $b(l)$\\
\ref{fig:biases} & 1 &  $\gamma_{\text{flash}}$, $\gamma_{\text{cloud}}$, $\gamma_{\text{land}}$\\
\ref{fig:leoper-leoper} & 5 &  $r_{\text{leonids}}(l)$, $r_{\text{perseids}}(l)$\\
\ref{fig:leoper-known-background} & 6 &  $r_{\text{base}}(l)$\\
\ref{fig:separate} & 3, 4 &  $r_{\text{base}}(l)$\\
\ref{fig:confidence}  & 7-12 & $r_{\text{base}}(l)$
\end{tabular}
\normalsize
\end{table}
\normalsize

\noindent We run the models described in Section \ref{sec:methods} with a range of different parameter settings. The set of models chosen allows us to investigate the differences in using human-vetted data and unvetted data, the effect of the confidence threshold, and the differences between GOES-16 and GOES-17.

\textbf{Model 1} is the reference model. It uses \texttt{machine} data with the standard threshold of 0.7 and only includes one bolide source, effectively capturing the total latitudinal variation, with background objects and meteor showers combined. \textbf{Model 2} uses the \texttt{human} data, and thus includes the stereo bias term $\gamma_{\text{stereo}}$ in Equation \ref{eq:b}, but is otherwise comparable to Model 1. It is intended to both estimate the stereo bias and test the robustness of the results. \textbf{Models 3-4} are the same as Model 1, but the data from GOES-16 and GOES-17 are not pooled, and a model is fit separately for each satellite. Models 3-4 are also intended to test the robustness of the results. \textbf{Model 5} differs from Model 1 in that it includes the Leonids and Perseids as bolide sources, hence leaving the rest as something closer to the background or sporadic source. Effectively, for any time $t$ five days around the solar longitude at which the Leonids peak \citep{showers}, the Leonids are included in $S(t)$ in Equation \ref{eq:R}, and likewise for the Perseids. We choose the Leonids and Perseids as these are quite prominent in GLM data (see Figure \ref{fig:sollon}) and have a latitudinal variation considerably different from that in Model 1. The choice of five days is based on visual inspection of the number of days around shower peaks in which GLM sees an increased rate (Figure \ref{fig:sollon}). Model 5 is designed to reveal whether or not GLM can recover the latitudinal variation of meteor showers, and hence the latitudinal variation of the Perseids and Leonids is treated as unknown and is estimated via Equation \ref{eq:r}. \textbf{Model 6} is similar to Model 5, but the latitudinal variation of the Leonids and Perseids is treated as known, and hence only a scaling factor is inferred via Equation \ref{eq:r-known}. Model 6 is designed to reveal how taking prominent meteor showers into account affects the estimates for the background rate. \textbf{Models 7-12} differ from Model 1 only in their different choice of confidence thresholds and test how sensitive results are to this choice. Table \ref{tab:models} summarizes the different models.

\begin{figure}
    \centering
    Latitudinal flux variation in Model 1 (unvetted data)

    \includegraphics[width=0.49\textwidth]{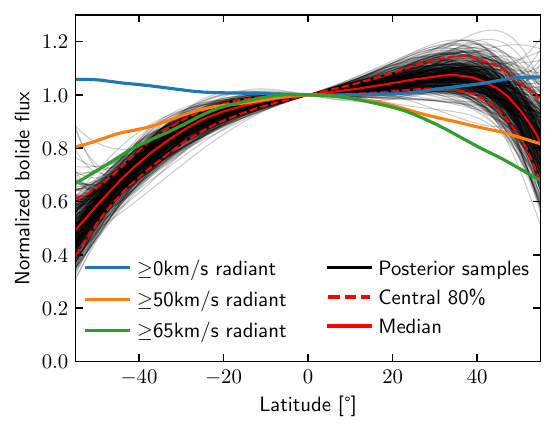}
    \includegraphics[width=0.49\textwidth]{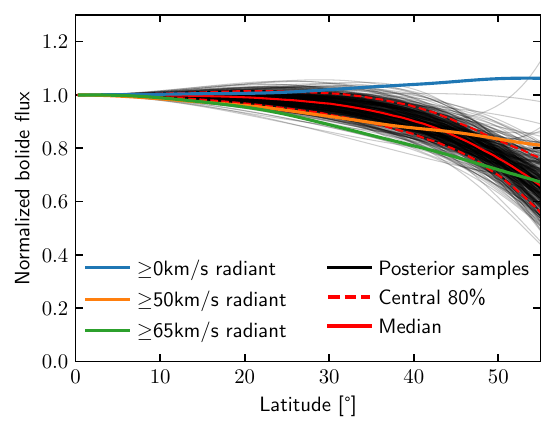}
    \caption{Bolide flux as a function of latitude in Model 1, normalized to 1 at $0^\circ$ latitude. The figure on the right folds latitude over the equator, so -50$^\circ$ becomes 50$^\circ$. Flux is given per unit area per unit time. Theoretical distributions obtained using the method of \cite{darrel} as described in Section \ref{sec:theoretical-flux}. In the plot on the left, we see that GLM bolide detections seem biased towards the North. In the plot on the right, while the observed curve does not precisely match any of the theoretical curves, the drop-off at latitudes distant from the equator is consistent with GLM being biased towards very fast objects, as according to the theoretical distributions plotted, there is a considerable drop-off in flux of such fast objects at higher latitudes.}\label{fig:machine-lat}
\end{figure}
\begin{figure}
    \centering
    Latitudinal flux variation and stereo bias in Model 2 (human-vetted data)

    \includegraphics[width=0.49\textwidth]{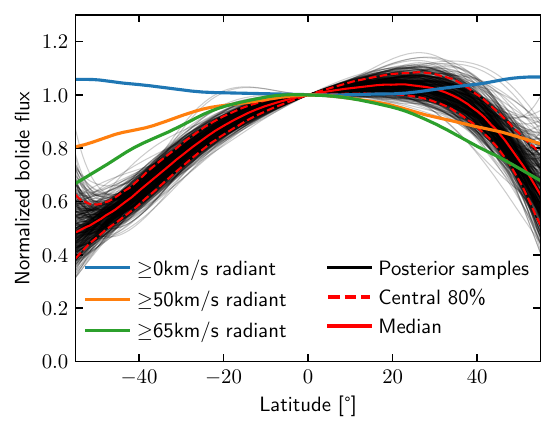}
    \includegraphics[width=0.49\textwidth]{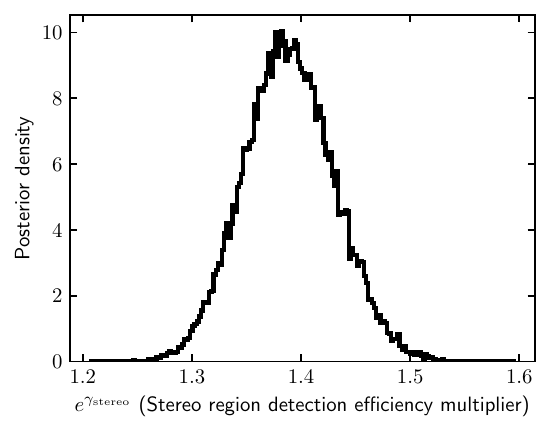}
    \caption{Left: Bolide flux in Model 2 as a function of latitude, normalized to 1 at $0^\circ$ latitude. The result is quite similar to Model 1. Right: The posterior distribution of $e^{\gamma_{\text{stereo}}}$ in Model 2. $\gamma_{\text{stereo}}$ is the coefficient applied to whether or not a bolide is in the stereo region in Equation \ref{eq:b}, and due to the model formulation $e^{\gamma_{\text{stereo}}}$ is the effective multiplier to detection efficiency in the stereo region.}\label{fig:human-lat}
\end{figure}
\begin{figure*}
    \includegraphics[width=\textwidth]{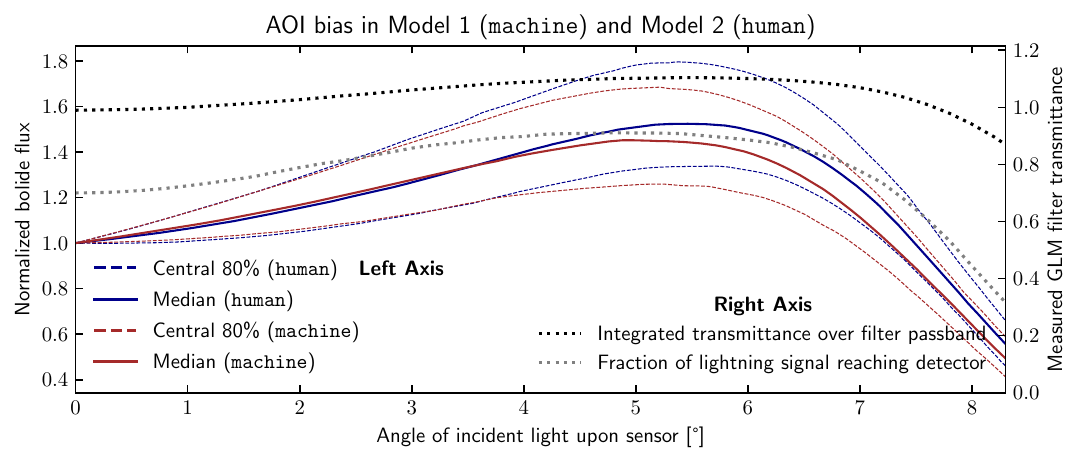}
    \caption{GLM bolide flux (left axis) as a function of the angle of incident light upon the sensor, along with curves of data representing GLM's sensitivity to light from lightning and the background (right axis, see also Figure \ref{fig:aoi}). The normalized bolide flux (expressed per unit area per unit time) is obtained from the posterior distributions of Model 1 and Model 2. The curves for both models are similar and roughly follow both curves of GLM transmittance data (especially that representing the fraction of lightning signal reaching the detector), peaking at about 5$^\circ$. Both curves of sensor data were measured from the GLM filter system \citep{glmfilter}.}
    \label{fig:fov}
\end{figure*}

\begin{figure}
    \centering
    Recovering the latitudinal flux variation of meteor showers in Model 5

    \includegraphics[width=0.50\textwidth]{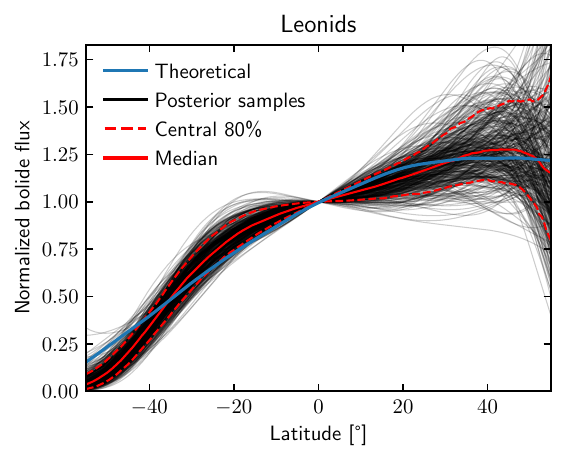}
    \includegraphics[width=0.49\textwidth]{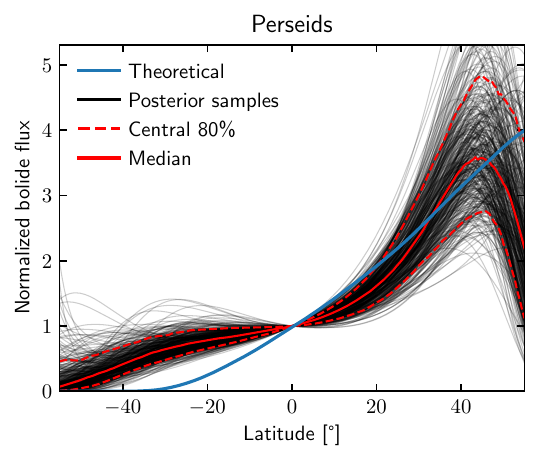}
    \caption{Meteor shower bolide flux (per unit area per unit time) as a function of latitude, normalized to 1 at $0^\circ$ latitude. More precisely, these are the functions $r_{\text{LEO}}$ and $r_{\text{PER}}$ in Equation \ref{eq:r} with the intercept set to zero to normalize them. These represent the latitudinal variation of all additional bolide flux which GLM sees during these meteor showers. Theoretical distributions obtained using the method of \cite{darrel} as described in Section \ref{sec:theoretical-flux} with radiant velocity, radiant declination, and solar longitude fixed to median values for the Leonids and Perseids \citep{showers}. We see that GLM does measure the fact that both meteor showers come from the Northern sky.}\label{fig:leoper-leoper}
\end{figure}

\begin{figure}
    \centering
    Latitudinal flux variation of background in Model 6

    \includegraphics[width=0.49\textwidth]{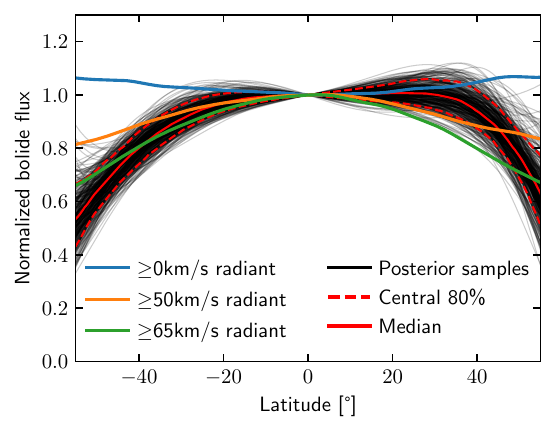}
    \includegraphics[width=0.49\textwidth]{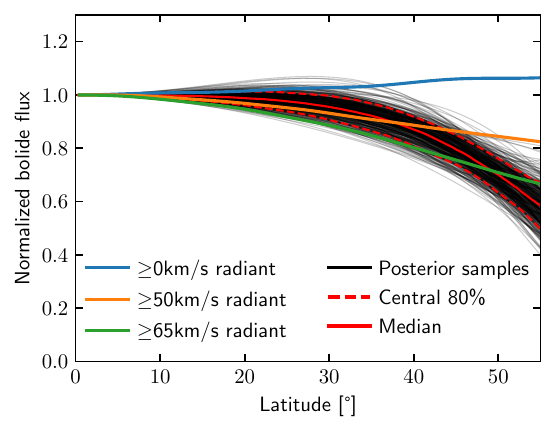}
    \caption{Background bolide flux (per unit area per unit time) as a function of latitude, normalized to 1 at $0^\circ$ latitude. Here ``Background'' is only with the Leonids and Perseids accounted for. The figure on the right folds latitude over the equator, so -50$^\circ$ becomes 50$^\circ$. Rate is per unit area per unit time. Theoretical distributions obtained using the method of \cite{darrel} as described in Section \ref{sec:theoretical-flux}. Much of the bias towards the North observed in Model 1 is no longer present in this model because the Leonids and Perseids, two Northern-sky meteor showers prominent in GLM data, are accounted for.}\label{fig:leoper-known-background}
\end{figure}

\begin{figure}
    \centering
    Models 3 and 4: separate fits for GOES-16 and GOES-17.

    \includegraphics[width=0.49\textwidth]{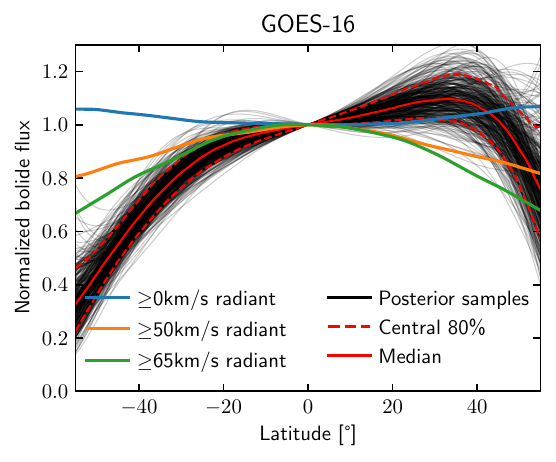}
    \includegraphics[width=0.49\textwidth]{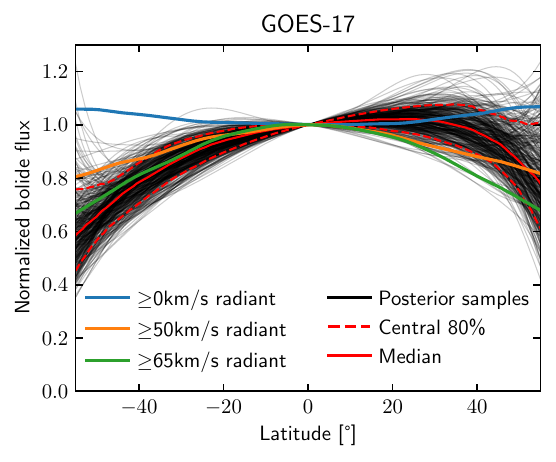}
    \caption{Bolide rate as a function of latitude, fitted separately for GOES-16 and GOES-17 (rather than combining the detections from both). Rate is per unit area per unit time. Theoretical distributions obtained using the method of \cite{darrel} as described in Section \ref{sec:theoretical-flux}. The distributions are similar, especially given the uncertainties. There doesn't appear to be any aspect of either distribution which is inconsistent with the other.}
    \label{fig:separate}
\end{figure}
\begin{figure*}
    \centering
    Background bolide flux in the \texttt{machine} database at different confidence thresholds.
    \includegraphics[width=0.49\textwidth]{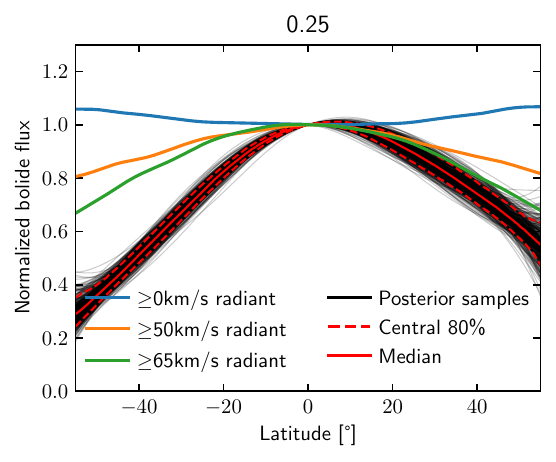}
    \includegraphics[width=0.49\textwidth]{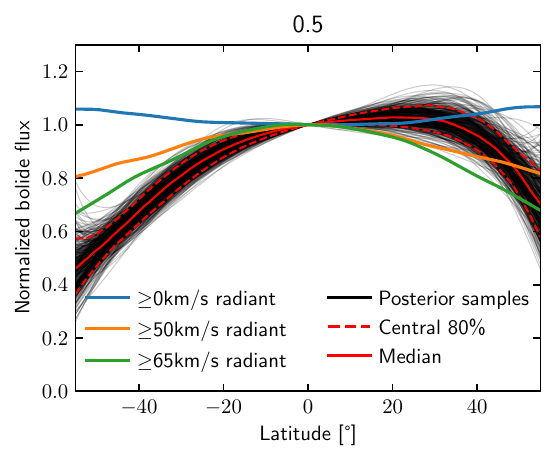}
    \includegraphics[width=0.49\textwidth]{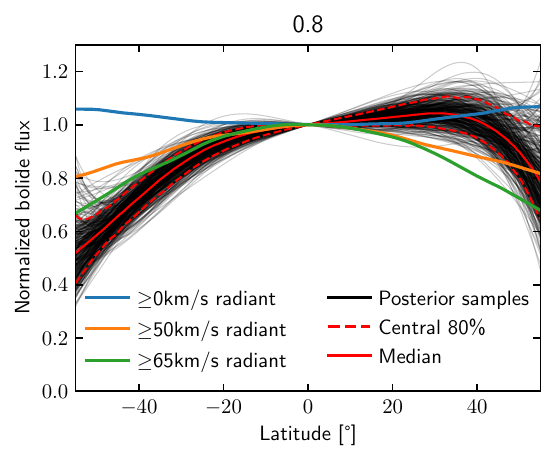}
    \includegraphics[width=0.49\textwidth]{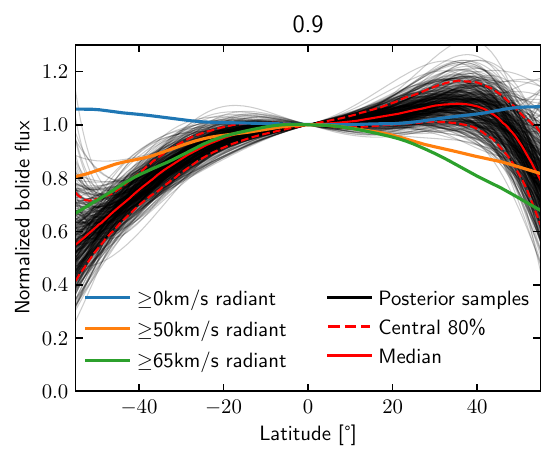}
    \includegraphics[width=0.49\textwidth]{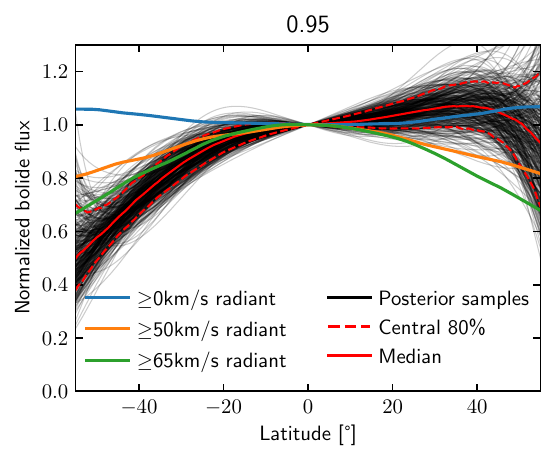}
    \includegraphics[width=0.49\textwidth]{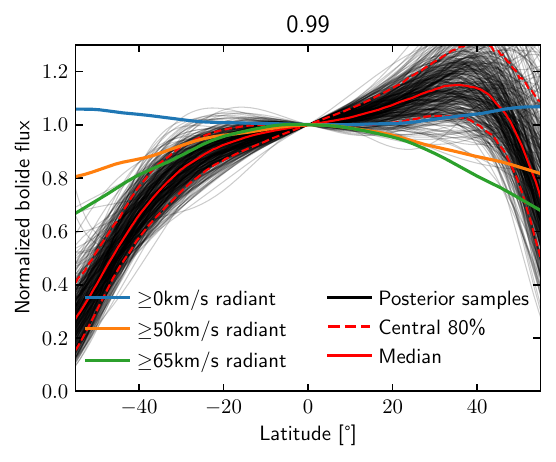}
    \caption{Background bolide flux as a function of latitude, computed by obtaining the posterior under the \texttt{machine} data at several classifier confidence thresholds. Flux is per unit area per unit time. Theoretical distributions obtained using the method of \cite{darrel} as described in Section \ref{sec:theoretical-flux}. Higher thresholds entail a wider posterior, as there are fewer detections. Lower thresholds emphasize the equator more, which may be due to false positives from solar glint and lightning activity. The results are robust at a wide range of confidence thresholds.}
    \label{fig:confidence}
\end{figure*}

For each model, after running the No-U-Turn sampler to obtain the posterior distributions, we can separately examine the posterior distributions of the polynomials for $b(d)$ and $r_s(l)$. Table \ref{tab:figures} summarizes the models and estimated variables studied in each figure.

Figure \ref{fig:machine-lat} shows the variation of the bolide rate in Model 1 as a function of latitude, with a decrease at latitudes distant from the equator and a bias towards the North. Figure \ref{fig:human-lat} shows similar results for Model 2, as well as measurements of $\gamma_{\text{stereo}}$, the stereo bias term. Figure \ref{fig:fov} displays the variation of the bolide rate as a function of the angle of incidence in Models 1 and 2. The alignment of the posterior bolide detection rate curves with the known efficiency of the sensor suggests that the model is recovering a true bias based on band shift due to incidence angle.

Figure \ref{fig:leoper-known-background} displays the variation of the non-Leonid or Perseid bolide rate in Model 6 as a function of latitude. The steep drop-off at latitudes far from the equator is consistent with GLM being strongly biased towards detecting faster objects. The bias towards the North is largely eliminated in this model compared to Model 1 (Figure \ref{fig:machine-lat}).

There are several ways to check the robustness of the model. Figure \ref{fig:leoper-leoper} displays the variation of the Leonid and Perseid bolide rate in Model 5 as a function of latitude. The inferred distributions reflect the fact that the Leonids and Perseids have radiants in the Northern sky and match theoretical predictions quite well. Figure \ref{fig:leoper-known-background} shows the latitudinal variation of the background bolide rate in Model 6, demonstrating how taking the Leonids and Perseids into account largely eliminates the bias towards the North. Figure \ref{fig:separate} shows the latitudinal variation when bolides detected by GOES-16 and GOES-17 are considered separately in Models 3 and 4, showing similar results. Figure \ref{fig:separate} also demonstrates the benefit of pooling the data from both satellites to obtain a tighter posterior that is less dependent on any possible effects of the geography in a satellite's FOV. Finally, Figure \ref{fig:confidence} shows the latitudinal variation when filtering the pipeline data at several different classifier confidence thresholds in Models 7-12 (not just 0.7 as used throughout), showing that it is similar for thresholds between 0.5 and 0.9.

\section{Discussion}

\subsection{Computed distribution}
\noindent A key finding, that GLM does indeed detect a decrease in bolide flux at latitudes far from the equator, is consistent with GLM being biased towards faster objects. This is because, according to the method in Section \ref{sec:theoretical-flux}, the flux of faster objects (on the order of 50-70 \unit{\km\per\s}) should decrease as one moves away from the equator. As discussed in Section \ref{sec:vel_ind}, GLM appears disproportionately sensitive to showers like the Leonids, which have a radiant velocity of \qty{70}{\km\per\s}, suggesting that GLM does indeed have this bias. This is also consistent with the observed bolide detections by solar hour depicted in Figure \ref{fig:solar} in Section \ref{sec:vel_ind}, which shows an overabundance of detections at points in the direction of the Earth's motion even while there does not appear to be a large effect of sunlight on bolide detection efficiency. It would also make physical sense, as, with equal mass, faster objects both deposit more energy into the atmosphere and produce more emissions at \qty{777}{\nm} \citep{vojacek}. However, because most GLM detections lie outside of the stereo region and do not have associated velocity data, and most detections in the stereo region are single-pixel observations with unreliable velocity data, adjusting for velocity is not currently possible. A full analysis of GLM velocity calculations, their uncertainty, and velocity-derived detection biases remains to be done.

Figure \ref{fig:leoper-leoper}, showing Leonid and Perseid bolide rates across latitudes, shows good agreement between the distributions detected by GLM and the theoretical distributions simulated using known properties of the showers. There are some discrepancies, though. The simplest to explain is that the Perseids have a higher-than-expected inferred flux in the South. This could be due to the presence of sporadic bolides above the base rate around the peak of the Perseids which would on average pull up the Southern tail of the model fit. Moreover, due to the model’s use of an exponential link function in Equation \ref{eq:r-known}, it is not easy for the model to infer a flux very close to zero as this would require the parameters to attain large negative values.

The other discrepancies in Figure \ref{fig:leoper-leoper} are harder to explain: the Leonids have a lower-than-expected inferred flux at the very Southern edge of the FOV and the Perseids seem to have a decreasing inferred flux North of $+40^\circ$. There may be additional, unaccounted-for factors affecting GLM's bolide detection efficiency at extreme latitudes. It is also possible that, though the model applies the same bias due to angle of incidence (and other factors) to all bolide sources, the biases are really more pronounced or less pronounced for these meteor showers. This could be due to them having a different brightness distribution, affecting how many objects drop below the level of detectability at a high angle of incidence. For the bias due to angle of incidence in particular, we can expect that a larger proportion of GLM-received light for Leonids and Perseids will come from oxygen line emissions due to their high speeds. This would cause the angle of incidence and corresponding band shift to have a greater effect on the fraction of their light reaching the detectors. Another possible explanation for these discrepancies observed is that the model effectively considers any excess detections around the peaks of the Leonids and Perseids as coming from these showers. If there are other meteor showers at play, or there are changes in the background flux during the showers, the inferred latitudinal variation for these showers will be affected.

Finally, Figure \ref{fig:leoper-known-background} shows that the observed excess of detections in the North in GLM data entirely disappears when taking the Leonids and Perseids into account. It seems that it is solely a result of GLM's sensitivity to meteor showers with high radiant velocities which by chance have Northern radiants.

\subsection{Estimated detection biases}
\begin{figure*}
    \centering
    \includegraphics[width=0.32\textwidth]{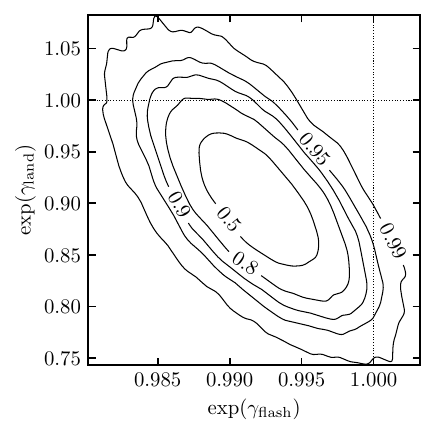}
    \includegraphics[width=0.32\textwidth]{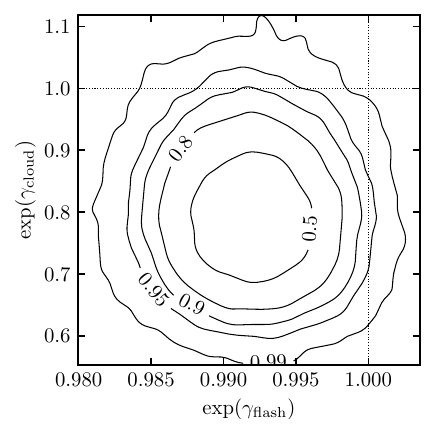}
    \includegraphics[width=0.32\textwidth]{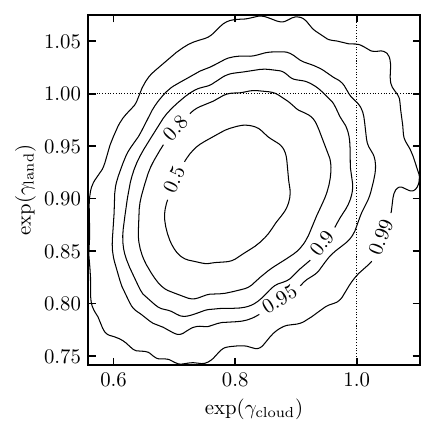}
    \caption{Joint posterior densities of $e^{\gamma_{\text{flash}}}$, $e^{\gamma_{\text{land}}}$, and $e^{\gamma_{\text{cloud}}}$ in Model 1. Respectively, these are: the multiplier on detection efficiency per unit of lightning flash density (flashes \unit{km^{-2}.year^{-1}}), the multiplier on detection efficiency over land, and the multiplier on detection efficiency for areas with constant cloud cover. The posterior density is represented by contours showing highest-posterior-density credible intervals.}
    \label{fig:biases}
\end{figure*}

\noindent Together, the models encompass five different possible detection biases. These are due to the angle of incidence, whether or not a detection is made over land, how much cloud cover a location has, how common lightning is at a location, and, for a human vetter, differing detection efficiency inside and outside of the stereo region.

Figure \ref{fig:fov}, which displays bolide rate as a function of the angle of incidence of the light entering the sensor, shows a strong dependence. This dependence aligns quite well with curves representing the measured transmittance of the filters in GLM, particularly the curve showing the fraction of lightning signal which reaches the detector. In the GLM passband, the light from lightning can be assumed to be entirely made up of oxygen emissions. But bolides are not like lightning, and GLM detects a combination of continuum radiation and oxygen line emissions during bolide impacts \citep{ogglm,jeff} and angle of incidence affects both the total efficiency of light transmission through the set of filters and shifts the passband of the narrowband filter nominally at \qty{777.4}{\nm}. Another effect may be a changing distribution of viewing geometries at different angles of incidence: while GLM can still detect bolides which don't move across the image plane, different viewing geometries can cause GLM to see more or less of the bolide head or tail. The normalized bolide flux curves in Figure \ref{fig:fov} also show greater variation across angles of incidence than the transmittance curves. This could be because a reduction in detected light leads to a reduced signal-to-noise ratio and, as most bolides are low-amplitude, an amplified reduction in the number of detectable bolides.

Thus the detection efficiency of bolides as a function of angle of incidence is a combination of different optical effects affecting both continuum and line emission and depends on bolide entry angle, velocity, mass, and composition. In turn, the observed bolide rate as a function of angle of incidence depends on the population of bolides. Any differences between the Leonids, Perseids, and the general population of GLM-detected bolides could explain the discrepancies in Figure \ref{fig:leoper-leoper}. Fully disentangling and understanding these effects is the subject of future work.

The other detection biases studied can each be encompassed in a single estimated coefficient. Figure \ref{fig:biases} shows estimates for the change in detection efficiency over land, in areas with lots of cloud cover, and per unit flash density (lightning flashes \unit{km^{-2}.year^{-1}}). We see that the coefficients on flash density and land are negatively correlated, which makes sense since lightning flash density is, on average, higher over land. It is also apparent that detection efficiency very likely decreases in areas with high lightning flash density. This is to be expected, as the classifier penalizes areas which would otherwise have too many false positives.

Independent of this flash density effect, it seems that detection efficiency is reduced over land, and likewise for areas with lots of cloud cover. GLM bolide detections are ultimately based on data downlinked from the GOES satellites which must be brighter enough than the background to pass an absolute noise threshold. This noise threshold increases when the background is brighter, as the sensor sees increased noise \citep{ogglm}. So our results are sensible, as both land and cloud cover tend to show up brightly in GLM background images. It is interesting that solar illumination doesn't appear to be considerably related to bolide detection rates in Figure \ref{fig:solar}, although that analysis doesn't adjust for any detection biases. In any case, the effect of a bright background doesn't appear to be as prominent in the \texttt{machine} database as might be expected from the analysis in \cite{ogglm}. This might be because bolides detected by the algorithm described in \cite{jeff} are not just a single threshold-crossing data point (as in the analysis in \cite{ogglm}), but typically contain many points which have energies considerably above the threshold. For such detections, a small increase in the threshold will generally only remove a few of the first and last dim points in the detection, and not eliminate it entirely. Future work ought to look closer at the effect of background brightness, perhaps by directly including averaged GLM background brightness in a model like that presented in Section \ref{sec:methods}.

\subsection{Comparison to previous empirical calculations}

\noindent \cite{darrel} and \cite{evatt} include empirical calculations for USG data. They both demonstrate a flux that appears to increase then decline as one moves away from the equator. In the USG data, the large error bars may indicate that the initial rise might be due to sampling error. However, the decline at high latitudes appears to be significant in both studies. In this work, a decline at high latitudes in USG data is hinted at in Figure \ref{fig:solar}, where there appear to be fewer detections than expected from theoretical models at low apparent solar altitudes (common at the poles). A decline at high latitudes is consistent with our findings in GLM data, though as \cite{darrel} note, the USG data does not seem to show a decline until latitudes that are outside the GLM field of view. But in principle USG data cannot be used for such studies as the detection efficiency of the sensors cannot be assumed to be uniform across the surface of the Earth. There is insufficient public information about USG sensor characteristics and their detection pipeline to debias USG data using a model similar to that used in this paper. USG data is, of course, useful for light curves and studying individual events, though care must still be taken as the detailed properties of the sensors are classified.

\cite{darrel} also include empirical calculations for GLM data. They again show an increase followed by a decline as one moves away from the equator, though the decline is much faster than in USG data. Our Figure $\ref{fig:fov}$ shows that detection rates peak at an angle of incidence which happens to correspond with about twenty degrees of latitude. Hence our Figure \ref{fig:machine-lat} does not show as much of an increase in flux as one moves away from the equator. Similarly, our results show that detection rates plummet at high angles of incidence of light into the sensor, and hence our results do not show as much of a decrease at high latitudes, though there is still a decline. In both cases, estimating and taking into account the bias due to the angle of incidence makes the variation observed across latitudes less extreme. We also eliminate the bias towards the stereo region---which Figure \ref{fig:human-lat} shows is quite large---by using pipeline data rather than human-vetted data, and use the computed FOV rather than the nominal FOV. For all of these reasons, our model represents an improvement on previous work.

In terms of our model structure, one additional advantage our model has is the parametric Equation \ref{eq:r} combined with the regularized horseshoe prior. We are able to offer much tighter distributions than completely nonparametric methods like binning and using Poisson errors which do not take full advantage of the data and lead to larger errors than necessary. It is reasonable to assume a parametric model with certain sparsity constraints, as the variation over latitudes should be smooth and not jump wildly from one bin to the next.

\subsection{Comparison to previous theoretical calculations}

\noindent Several previous studies have made theoretical calculations of the latitudinal distribution of impacts. These studies fall into two categories:
\begin{enumerate}[label=(\arabic*)]
    \item \cite{halliday} and \cite{evatt} proceed from the assumption that the vast majority of impactors come from the ecliptic, and with some calculations obtain the resulting distributions for given velocities. The former simply reports these distributions, while the latter uses the velocity distribution in USG data to derive a final distribution.
    \item \cite{lefeuvre} and \cite{darrel} proceed from models of asteroids ejected from the main belt by interactions with Jupiter and their computed impact probabilities to produce distributions of ecliptic latitude and velocity, and run Monte-Carlo simulations to obtain final distributions. The latter uses updated NEO distributions and impact probabilities.
\end{enumerate}
The results of these studies can be characterized by their pole/equator impact flux ratio. The studies in (1) predict a large decrease in flux at high latitudes, with pole/equator ratios as low as 0.5 (depending on impactor velocity), while those in (2) predict either a roughly uniform distribution or an increase in flux at high latitudes, with pole/equator ratios from 0.96 in \cite{lefeuvre} to 1.08 in \cite{darrel} (see erratum, 2023). This difference between the two study categories is understandable, as including objects with high inclinations in the calculations will increase the relative flux at the poles. The assumption that the vast majority of impactors come from the ecliptic appears to be an oversimplification according to observations of NEOs and models of their collision probabilities \citep{darrel}. Measurements of smaller, sporadic meteors also show strong sources of such meteors which do not lie in the ecliptic plane \citep{sporadic-dist}.

Our results, which show a decrease in flux at high latitudes, \textit{cannot} be taken as supporting the studies in (1) over the studies in (2), due to the possible velocity bias in GLM detections which would in fact create such an observed drop-off at high latitudes. In fact, if GLM does indeed have an overwhelming bias towards objects with high radiant velocities, our results are consistent with our theoretical calculations using the methods of \cite{darrel} described in Section \ref{sec:theoretical-flux} (as seen in Figure \ref{fig:leoper-known-background}, among others). Other differences between high- and low-inclination sources like a different mass index could also lead to GLM detecting more bolides from low-inclination sources than might be expected from theoretical models and contribute to producing the observed distribution.

\subsection{Future work}
\noindent If the GLM detection rates for impactors of different masses and velocities are known, it should be possible to correct our estimates of latitudinal variation accordingly. Fortunately, several ground-based meteor networks have cameras in the GLM field of view, and their ground-based observations allow precise estimates of orbits, velocities, and some estimation of physical characteristics. Given that ground-based data contains much fainter objects than GLM data does, it may be possible to assume that ground-based data will detect all impacts above a certain mass and velocity (assuming good observing conditions with respect to sunlight and weather). It would then be possible to see what proportion of ground-based detections with certain characteristics and velocities are detected by GLM, and finally to completely debias GLM's observed distribution. However, there are challenges associated with this, including that ground-based data is affected by weather and, in particular, sunlight and clouds, which may skew the data in unexpected ways.

More work could also be done on fusing ground-based data with GLM data to improve models estimating velocity and physical characteristics from GLM data alone. Also, models of GLM sensitivity could be paired with simulations of bolide spectra as viewed from GLM's perspective. This would allow simulated GLM data to be created and injected into the detection pipeline, fully characterizing the detection efficiency for bolides with different properties and at different locations.

\section{Conclusion}

\noindent We have developed methods to debias GLM bolide detections in order to rigorously study detection rates. Our methods take into account the precise field of view of the GLM instruments and discretize it in space and time. Using a Poisson regression, we are able to simultaneously estimate the latitudinal variation of bolides detected by GLM and detection biases due to sensor properties. Using our model we have partially debiased the latitudinal variation of bolide flux detected by GOES GLM.

We have also presented numerous results that indicate that GLM experiences a strong detection bias towards high-velocity objects, and have demonstrated how this can generate our observed partially-debiased distribution. We cannot yet conclusively show the existence of this bias or measure it. Nevertheless, future works should be aware that the population of objects detected by GLM---which we think is dominated by high-velocity impactors on retrograde orbits, including several meteor showers---may not be what is anticipated.

Our method is able to closely recover the well-understood latitudinal distributions of meteor showers like the Leonids and Perseids. Yet we advise against using the results of this work in confirming or rejecting theoretical models of overall impact distributions, as GLM's bolide detection biases are still not understood well enough to obtain a completely debiased estimate. We are unable to recover an overall debiased distribution as velocities are not measured accurately enough, and bolide velocity is related to both its radiant and, presumably, its probability of detection by GLM.

However, with future work on more reliably estimating velocities from GLM data and fusing GLM data with other sources, it may be possible to recover from the presumed velocity bias, and the models and methods presented here can be extended to take velocity into account. Any future works performing statistical analyses of GLM detections may be affected by a bias due to angle of incidence or other detection biases and could benefit by taking these into account using the methods presented here or others.

\section*{Data \& Code availability}
\noindent All code involved in this work and the bolide data used are available at \url{https://github.com/anthonyozerov/bolide-stats}. This repository also contains intermediate results from Section \ref{sec:theoretical-flux} that may be of interest, such as estimated latitude distributions for different velocities and meteor showers  as well as the 3D ecliptic latitude, Sun-centered ecliptic longitude, and radiant velocity probability distribution. Readers may refer to \cite{granvik} for the sample of 802,000 synthetic NEOs, to \cite{flashdensity} for the lightning flash density data used, and to \cite{modis} for the cloud fraction data used. The corresponding author will also provide these upon request, and code for reading these data is available at the repository above.

\section*{CRediT authorship contribution statement}
\noindent \textbf{Anthony Ozerov}: Conceptualization, Original Draft, Methodology, Software. \textbf{Jeffrey C. Smith}: Conceptualization, Supervision. \textbf{Jessie L. Dotson}: Project management, Review \& Editing. \textbf{Randolph S. Longenbaugh}: Data Curation. \textbf{Robert L. Morris}: Review \& Editing.

\section*{Acknowledgments}
\noindent Darrel Robertson provided code to calculate theoretical distributions across latitudes and much helpful discussion. Katrina Virts provided the data on the GLM field of view, without which a proper analysis would not have been possible. We are grateful to two peer reviewers for their expertise and feedback. We would also like to thank developers of open-source software, including: Python, the scientific Python stack, Astropy, PyMC, and NumPyro. Work carried out for this project is supported by NASA’s Planetary Defense Coordination Office (PDCO). Resources supporting this work were provided by the NASA High-End Computing (HEC) Program through the NASA Advanced Supercomputing (NAS) Division at Ames Research Center. AO, JS, and RM are supported through NASA Cooperative Agreement 80NSSC19M008.

\bibliography{references}

\end{document}